\documentclass[%
preprint,
bibnotes,
 amsmath,amssymb,
 aps,
]{revtex4-1}

\usepackage[pdftex]{graphicx}% Include figure files
\usepackage{dcolumn}% Align table columns on decimal point
\usepackage{bm}% bold math
\usepackage{theorem}
\theorembodyfont{\normalfont}

\usepackage{color}
\usepackage{comment}
\usepackage{txfonts}
\usepackage{algorithm}
\usepackage{algorithmic}
\usepackage{bbm}
\usepackage{mathrsfs}

\newcommand{\argmin}{\mathop{\rm arg~min}\limits}

\newcounter{num}

\newcommand{\rnum}[1]{\setcounter{num}{#1}\roman{num}}

\newif\iffigure
\figuretrue

\begin{document}

\preprint{APS/123-QED}

\title{Interpretable Conservation Law Estimation\\ by Deriving the Symmetries of Dynamics from Trained\\ Deep Neural Networks}% Force line breaks with \\

\author{Yoh-ichi Mototake}
\email{mototake@ism.ac.jp}
%\email{okada@edu.k.u-tokyo.ac.jp}
% \altaffiliation[Also at ]{Department of art and science, University of Tokyo.}%Lines break automatically or can be forced with \\
%\author{Takashi Ikegami}%
\affiliation{%
The Institute of Statistical Mathematics, Tachikawa, Tokyo 190-8562, Japan\\
}%


%\affiliation{
%\affiliation{
%\affiliation{%


\date{\today}% It is always \today, today,

\begin{abstract}
%Understanding complex systems with their reduced model is one of the central roles in scientific activities. Although several machine learning methods, such as deep neural networks (DNNs), can be used to infer a reduced model, these methods have the disadvantage of limited interpretability. 
%We propose a new approach to extracting interpretable physical laws from DNNs trained with physical data. 
%The approach does not infer the reduced model directly but provides interpretable conservation laws to physicists to build an interpretable reduced model. 
%The approach is realized using two newly developed methods. One is a method of estimating transformations that make the target system invariant. The other is a method of inferring hidden conservation laws that use Noether's theorem. 
%These methods are developed by deriving the relationship between a manifold structure of time-series dataset and the necessary conditions for Noether's theorem. 
%The feasibility of the approach has been verified in some primitive cases for which the conservation law is well known. 
%We also apply the approach to conservation law estimation for a more practical case that is a large-scale collective motion system in the metastable state, and we obtain a result consistent with that of a previous study. 
Understanding complex systems with their reduced model is one of the central roles in scientific activities.  Although physics has greatly been developed with the physical insights of physicists, it is sometimes challenging to build a reduced model of such complex systems on the basis of insights alone. We propose a novel framework that can infer the hidden conservation laws of a complex system from deep neural networks (DNNs) that have been trained with physical data of the system. The purpose of the proposed framework is not to analyze physical data with deep learning, but to extract interpretable physical information from trained DNNs. With Noether's theorem and by an efficient sampling method, the proposed framework infers conservation laws by extracting symmetries of dynamics from trained DNNs. The proposed framework is developed by deriving the relationship between a manifold structure of time-series dataset and the necessary conditions for Noether's theorem. 
The feasibility of the proposed framework has been verified in some primitive cases for which the conservation law is well known. 
We also apply the proposed framework to conservation law estimation for a more practical case that is a large-scale collective motion system in the metastable state, and we obtain a result consistent with that of a previous study. 
\end{abstract}

\pacs{Valid PACS appear here}% PACS, the Physics and Astronomy
\maketitle

%\tableofcontents

%%% Keywords are not needed any longer. %%%
%%%\kword{keyword1, keyword2, keyword3, \ldots}
%%%

\section{Introduction}
Understanding complex systems with their reduced models is one of the central roles in scientific activities. 
Some complex systems are modeled as low-dimensional canonical dynamical systems. 
For example, reduced models have been developed for large-scale collective motion systems, which are a type of large-scale complex system with order (e.g., plasma, acoustic waves, or vortex systems)~\cite{Tomonaga:1950zz, bohm1951collective, pines1952collective, tomonaga1955elementary, saffman1992vortex}. 
To develop reduced models, collective coordinates, such as the Fourier basis of a density or charge distribution~\cite{Tomonaga:1950zz, bohm1951collective, pines1952collective, tomonaga1955elementary}, or a vortex feature space~\cite{saffman1992vortex}, have been introduced. 
Then, a Hamiltonian that describes the coarse-grained properties of a dynamical system has been derived. 
Thus, to develop a reduced model, it is necessary to introduce collective coordinates and derive the Hamiltonian in the coordinates. 
The obtained Hamiltonian is verified by confirming that it can reconstruct the properties of the phenomena analyzed. 
This approach relies heavily on the physical insights of physicists; it would not work to model a dynamical system that features a more complicated structure. One example is the collective motion of living things such as fish or birds; such systems frequently have stable but very complicated patterns in a metastable state~\cite{vicsek2012collective, ikegami2017life}.\par
The problem we consider here is how to infer the reduced model using machine learning methods. 
As mentioned above, this involves the solution of two problems: estimation of a coordinate system and construction of a reduced model in the coordinate system. 
One way to solve these problems is to construct a Hamiltonian based on a given coordinate system and search for a coordinate system that improves the model. 
Several machine learning methods for inferring the Hamiltonian from a time-series dataset have been developed~\cite{schmidt2009distilling, NIPS2019_9672, toth2019hamiltonian, bondesan2019learning}. 
These methods can be broadly divided into two types. 
In one type, the Hamiltonian is inferred by regressing the data with an explicit function, such as the linear sum of multiple basis functions~\cite{schmidt2009distilling}. 
However, in the case of inferring a reduced model that consists of complicated unknown basis functions, the method only infers the approximated reduced model using an approximated function, such as a polynomial function.  
In the second type, a Hamiltonian is modeled by a deep learning technique~\cite{NIPS2019_9672, toth2019hamiltonian, bondesan2019learning}. In this case, an explicit function used in the first one is not required. 
On the basis of these machine learning methods, the search for the coordinate system could be performed using statistical criteria such as the prediction or generalization error of the inferred Hamiltonian.\par
There are inherent difficulties in building a reduced model using the machine learning approach. 
Such an approach finds a Hamiltonian that has properties that only hold for the given data. 
Historically, physicists have achieved great success in constructing reduced models by abstracting knowledge obtained from observational data and building universal models that can explain various physical phenomena, not just the given data. 
For example, in thermodynamics, a reduced model that describes the molecular motion of a gas was linked to chemical reaction theory by Gibbs~\cite{gibbs1876, gibbs1878}. This is one of the most successful uses of a reduced model. 
That is, a good reduced model and a good coordinate system mean that the performance is high not only for the given data.\par
To reach such a successful reduced model, it is important to interpret the knowledge obtained during data analysis and develop a model that can be applied to different phenomena by combining the explicit and implicit knowledge of physics. 
In general, an inferred Hamiltonian modeled by deep neural networks (DNNs) is hardly interpretable, because DNNs are models with enormous degrees of freedom. 
If all physical knowledge is quantified, it will be possible to construct a reduced model with a DNN, but this is an impractical assumption at present. 
Therefore, it is difficult by a machine learning approach to realize the same function as a physicist, who can flexibly interpret phenomena by utilizing explicit or implicit physical knowledge and construct a reduced model. \par
To overcome this problem, we attempt to extract abstract information directly from physical data without constructing a reduced model. 
A coordinate system can be selected on the basis of the information. 
Furthermore, the obtained information can also help physicists construct a reduced model. 
The purpose of this study is to develop a machine learning framework that extracts interpretable abstract information from physical data and assist physicists in building reduced models.\par
The proposed method is developed using knowledge about DNNs. Results of several studies~\cite{Irie_1990,Hinton_Reducing_2006,brahma2016deep,basri2016efficient,Rifai_The_2011,mototakearob} suggest that DNNs can model the distribution of datasets as manifolds, which can be embedded in a low-dimensional Euclidean space. 
Studies applying DNNs to physical data have employed a time-series dataset from the phase space (comprising position and momentum)~\cite{yeo2017model,morton2018deep,rudy2018deep,takeishi2017learning,lusch2018deep} or a spin system dataset from the configuration space~\cite{ohtsuki2016deep,ohtsuki2017deep,broecker2017machine,ch2017machine,carrasquilla2017machine,tanaka2017detection,saito2017machine,van2017learning,zhang2018machine}. 
In such datasets, the manifold structure, which implies that the system has a small degree of freedom, can be constructed by considering certain physical constraints, such as a conservation law. 
That is, a manifold structure modeled by a DNN can represent the conservation law or order of the system.\par
The proposed method is derived from Noether's theorem~\cite{noether1918nachr}, which connects the symmetry of the Hamiltonian and the conservation law. 
We derive the relationship between the symmetry of the Hamiltonian system and the distribution of the time-series dataset of a dynamical system. 
On this basis, we develop a method of inferring the symmetry of a data manifold modeled by a deep autoencoder~\cite{Hinton_Reducing_2006} and determine conservation laws of the system. 
To infer the conservation laws, we only need the tangent space of the manifold of the continuous transformation group that corresponds to the symmetry of the system. 
Therefore, unlike Hamiltonian estimation, conservation law estimation only requires manifold modeling with at most first-order accuracy. This means that the conservation law can be inferred with arbitrary precision by polynomial approximation.
\par
This paper is organized as follows. 
In Sec.~\ref{sec_noether}, we show the derivation of the relationship between the symmetry of the time-series dataset distribution and the conservation law using Noether's theorem. 
In Sec.~\ref{sec_extract}, we describe our proposed method of inferring the symmetry of the time-series data manifold. 
In Sec.~\ref{sec_consest}, we also describe another proposed method of inferring the conservation law from the obtained symmetry. 
In Sec.~\ref{results}, to confirm the effectiveness of the proposed methods, we apply them to three cases, one T(1) and two SO(2) systems, corresponding to constant-velocity linear motion, a central force system, and a large-scale collective motion system called the Reynolds model~\cite{reynolds1987flocks}.  
In Sec.~\ref{summary}, we present a summary and discussion.

\section{Theory}
\label{theory}
\subsection{Noether's theorem}
\label{sec_noether}
%系の対称性と保存量との関係について，
%ネーターによって証明された定理である\cite{noether1918nachr}．
%ここで，d次元一般化座標を，
%${\bf q}(t)$と${\bf p}(t)$
%とし，系のハミルトニアンを$H(q,p;t)$とする．
Noether's theorem connects continuous symmetries of a Hamiltonian system with conservation laws~\cite{noether1918nachr}. 
It is often described in the $(2d+1)$-dimensional extended phase space $\Gamma \times \mathbb{R}, \: \left(\textit{\textbf{q}}, \textit{\textbf{p}}\right)\coloneqq\left(q_0=t,q_1,\cdots,q_d,p_1,\cdots,p_d\right)$. The theorem can also be described in the $(2d+2)$-dimensional space $\Gamma \times \mathbb{R} \times \mathbb{R}, \left(q_0=t,q_1,\cdots,q_d,p_0=-H,p_1,\cdots,p_d\right)$. In this study, we describe the theory in the $(2d+2)$-dimensional space as follows. 
We consider Hamiltonian systems in the $(2d+2)$-dimensional space $\Gamma \times \mathbb{R} \times \mathbb{R}$, and restrict ourselves to the case where the system's Hamiltonian belongs to a $C^2$ class function $H(\textit{\textbf{q}},\textit{\textbf{p}})$. 
The Hamiltonian representation of Noether's theorem is described as follows~\cite{struckmeier2002canonical}. 
%今，ハミルトニアン$H(q,p;t)$が，ある微小な座標変換
Assume that $H(\textit{\textbf{q}},\: \textit{\textbf{p}})$ and the canonical equations of motion  $\frac{\partial{H(\textit{\textbf{q}},\textit{\textbf{p}})}}{\partial{q_i}} = -\dot{p_i}$ and $\frac{\partial{H(\textit{\textbf{q}},\textit{\textbf{p}})}}{\partial{p_i}} = \dot{q_i}$ are invariant under the infinitesimal transformation $(q'_i, p'_i) = (q_i + \delta q_{ij}, p_i + \delta p_{ij})$, 
%（ただし，j=1 $\sim$ d）に対して普遍であるとする．
where $i=1, \dots, d$, and $j$ is the index of the direction of the infinitesimal transformation corresponding to a conservation law. 
%すると，ネーターの定理より
%微小変換の生成子$G_{\delta}$が
%時間によらず普遍な量となる．
Then, on the basis of Noether's theorem, the conserved value $G_j$ satisfies the following equation:
\begin{equation}
\label{noether}
(\delta q_{ij},\delta p_{ij}) = \left( \frac{\partial{G_{j}}}{\partial{p_i}}, -\frac{\partial{G_{j}}}{\partial{q_i}} \right).
\end{equation}
%つまり，保存量となる．
%ここで微小変換の生成子$G_{\delta}$は，
%\begin{equation}
%(\delta q_j,\delta p_j) = \left( \frac{\partial{G_{\delta}}}{\partial{q_j}},\frac{\partial{G_{\delta}}}{\partial{p_j}} \right)
%\end{equation}
%を満たすものとして定義される．
The canonical transformation that makes the Hamiltonian system invariant is given as 
\begin{eqnarray}
\mathbbm{c}_{\rm inv}(\boldsymbol{\theta}):\: \Gamma \times \mathbb{R} \times \mathbb{R} &\longrightarrow& \Gamma \times \mathbb{R} \times \mathbb{R},\\ 
\:(\textit{\textbf{q}},\textit{\textbf{p}}) &\longmapsto& (\mathbf{\mathcal{Q}},\mathbf{\mathcal{P}}) \coloneqq \textbf{(}\mathbf{\mathcal{Q}}(\textit{\textbf{q}},\textit{\textbf{p}}, \boldsymbol{\theta}),\mathbf{\mathcal{P}}(\textit{\textbf{q}},\textit{\textbf{p}}, \boldsymbol{\theta})\textbf{)},
\end{eqnarray}
where $\mathbf{\mathcal{Q}}(\textit{\textbf{q}},\textit{\textbf{p}}, \boldsymbol{\theta})$ and $\mathbf{\mathcal{P}}(\textit{\textbf{q}},\textit{\textbf{p}}, \boldsymbol{\theta})$ represent the invariant transformation functions of coordinate $(\textit{\textbf{q}},\textit{\textbf{p}})$ to $(\mathbf{\mathcal{Q}},\mathbf{\mathcal{P}})$, and $\boldsymbol{\theta}$ represents a $d_{\theta}$-dimensional continuous parameter characterizing transformation that satisfies  $\mathbf{\mathcal{Q}}\left(\textit{\textbf{q}},\textit{\textbf{p}},\boldsymbol{\theta} = \vec{0}\right) = \textit{\textbf{q}}$, and $\mathbf{\mathcal{P}}\left(\textit{\textbf{q}},\textit{\textbf{p}},\boldsymbol{\theta} = \vec{0}\right) = \textit{\textbf{p}}$. 
We call this transformation an invariant transformation in this paper. 
A set of the invariant transformations characterized by the continuous parameters $\boldsymbol{\theta}$ forms a Lie group. 
By the first-order Taylor expansion of $\mathcal{Q}_i(\textit{\textbf{q}},\textit{\textbf{p}}, \boldsymbol{\theta})$ and $\mathcal{P}_i(\textit{\textbf{q}},\textit{\textbf{p}}, \boldsymbol{\theta})$ around $\boldsymbol{\theta} = \vec{0}$, we have the infinitesimal transformation 
\begin{equation}
(\delta q_{ij},\delta p_{ij}) = \left(\varepsilon \left.\frac{\partial \mathcal{Q}_i(\textit{\textbf{q}},\textit{\textbf{p}}, \boldsymbol{\theta})}{\partial \theta_j}\right|_{\boldsymbol{\theta} = \vec{0}}, \varepsilon \left.\frac{\partial \mathcal{P}_i(\textit{\textbf{q}},\textit{\textbf{p}}, \boldsymbol{\theta})}{\partial \theta_j}\right|_{\boldsymbol{\theta} = \vec{0}}\right),
\end{equation}
where $|\varepsilon| \ll 1$.\par
Note that the dimension of continuous parameter $d_{\theta}$ corresponds to the number of conservation laws, and with our proposed methods, we estimate conservation laws including $d_{\theta}$.

\subsection{Invariance of Hamiltonian and time-series dataset}
%\subsection{Time series data of dynamical system and Noether's theorem}
%次に，ネーターの定理を適用するために必要な
%ハミルトニアン$H(q,p;t)$の微小変換に対する普遍性を，
%位相空間での時系列データ$\{{\bf q}^i(t),{\bf p}^i\}_{i=1}^N$
%から得る方法について述べる．
We show the relationship between such an invariant transformation and 
the time-series dataset of a dynamical system in the $(2d+2)$-dimensional space $(\textit{\textbf{q}},\textit{\textbf{p}})$. 
Here, we define the $N$ sample time-series dataset $D$ as $D \coloneqq \left\{\textit{\textbf{q}}_{t_i}^i, \textit{\textbf{p}}_{t_i}^i,\textit{\textbf{q}}_{t_i+\Delta t}^i, \textit{\textbf{p}}_{t_i+\Delta t}^i\right\}_{i=1}^N$, where $\textit{\textbf{q}}_{t_i}^i$ and  $\textit{\textbf{p}}_{t_i}^i$ represent the generalized position and momentum at time $t_i$, and $t_i + \Delta t$ represents a time evolution of $\Delta t$.\par
The transformation of the $(2d+2)$-dimensional space $(\textit{\textbf{q}},\textit{\textbf{p}})$ is defined as  
\begin{eqnarray}
\label{transeq}
\mathbbm{c}:\: \Gamma \times \mathbb{R} \times \mathbb{R} &\longrightarrow& \Gamma \times \mathbb{R} \times \mathbb{R},\\ 
\:(\textit{\textbf{q}},\textit{\textbf{p}}) &\longmapsto& (\mathbf{Q},\mathbf{P}) \coloneqq \textbf{(}\mathbf{Q}(\textit{\textbf{q}},\textit{\textbf{p}}),\mathbf{P}(\textit{\textbf{q}},\textit{\textbf{p}})\textbf{)},
\end{eqnarray}
where $\mathbf{Q}(\textit{\textbf{q}},\textit{\textbf{p}})$ and $\mathbf{P}(\textit{\textbf{q}},\textit{\textbf{p}})$ represent transformations functions of coordinate $(\textit{\textbf{q}},\textit{\textbf{p}})$ to $(\mathbf{Q},\mathbf{P})$; the transformation is not limited to the invariant transformation. 
It is assumed that $\mathbbm{c}$ has the inverse transformation
\begin{eqnarray}
\mathbbm{c}^{-1}:\: \Gamma \times \mathbb{R} \times \mathbb{R} &\longrightarrow& \Gamma \times \mathbb{R} \times \mathbb{R},\\
\:(\textit{\textbf{Q}},\textit{\textbf{P}}) &\longmapsto& (\textit{\textbf{q}},\textit{\textbf{p}}) \coloneqq \boldsymbol{(}\textit{\textbf{q}}(\textit{\textbf{Q}},\textit{\textbf{P}}),\textit{\textbf{p}}(\textit{\textbf{Q}},\textit{\textbf{P}})\boldsymbol{)}.
\end{eqnarray}
The transformed Hamiltonian $H'(\textit{\textbf{q}},\textit{\textbf{p}})$ obeying this transformation is defined as $H'(\textit{\textbf{Q}},\textit{\textbf{P}}) \coloneqq H\left(\textit{\textbf{q}}(\textit{\textbf{Q}},\textit{\textbf{P}}),\textit{\textbf{p}}(\textit{\textbf{Q}},\textit{\textbf{P}})\right)$. 
The necessary and sufficient condition for the transformation $\mathbbm{c}$ acting on 
$H(\textit{\textbf{q}},\textit{\textbf{p}})$ to be identical, $H'(\textit{\textbf{q}},\textit{\textbf{p}}) \equiv H(\textit{\textbf{q}},\textit{\textbf{p}})$, is equivalent to
\begin{eqnarray}
\forall{E},\: \{\textit{\textbf{q}},\textit{\textbf{p}}\:\mid \: H(\textit{\textbf{q}},\textit{\textbf{p}}) = E\}= \{\textit{\textbf{Q}},\textit{\textbf{P}}\:\mid \:
H(\textit{\textbf{q}},\textit{\textbf{p}}) = E\}.
\label{eq_hamiltonian_condition}
\end{eqnarray}
This condition is derived in Appendix~\ref{appendix_hinv} and implies that the transformation invariance of a Hamiltonian is equivalent to that of the energy surface at each energy level in the space $\Gamma \times \mathbb{R} \times \mathbb{R}$. If the time-series dataset $D$ has all possible data points under the Hamiltonian $H(\textit{\textbf{q}},\textit{\textbf{p}})$, the subset of $D$ with respect to $\textit{\textbf{q}}_{t_i}^i$ and  $\textit{\textbf{p}}_{t_i}^i$ is understood as this energy surface.\par
%と等価である．
%エネルギーを無限小の間隔で刻んみ，そのうちの一つを$E_i$とする．
%そして，
%同様に運動方程式を不変とする変換は，
\subsection{Invariance of canonical equations and time-series dataset}
Next, we consider the relationship between the invariance of canonical equations of motion and the time-series dataset of the dynamical system. 
If the canonical equations of motion are discretized with respect to time differentiation, the discretized canonical equations of motion are obtained as
\begin{eqnarray}
    {\textit{\textbf{q}}}_{t+\Delta t} &=&
    \textit{\textbf{u}}(\textit{\textbf{q}}_{t},\textit{\textbf{p}}_{t}) \coloneqq \frac{\partial{H(\textit{\textbf{q}}_{t},\textit{\textbf{p}}_{t})}}{\partial{\textit{\textbf{p}}_{t}}}\Delta t + \textit{\textbf{q}}_{t},\\
    {\textit{\textbf{p}}}_{t+\Delta t} &=& {\textit{\textbf{v}}}(\textit{\textbf{q}}_{t},\textit{\textbf{p}}_{t}) \coloneqq -\frac{\partial{H(\textit{\textbf{q}}_{t},\textit{\textbf{p}}_{t})}}{\partial{\textit{\textbf{q}}}_{t}}\Delta t + {\textit{\textbf{p}}}_{t},
\end{eqnarray}
where $\textit{\textbf{q}}_{t}$ and $\textit{\textbf{p}}_{t}$ represent the variables that evolved according to time $t$, and $\textit{\textbf{u}}(\textit{\textbf{q}}_{t},\textit{\textbf{p}}_{t})$ and $\textit{\textbf{v}}(\textit{\textbf{q}}_{t},\textit{\textbf{p}}_{t})$ are elements of the $C^1$ map $\mathbbm{u}$ defined as 
\begin{eqnarray}
\mathbbm{u}:\:\Gamma \times \mathbb{R} \times \mathbb{R} &\longrightarrow& \Gamma \times \mathbb{R} \times \mathbb{R},\\
\:(\textit{\textbf{q}}_{t}, \textit{\textbf{p}}_{t}) &\longmapsto& (\textit{\textbf{q}}_{t+\Delta t},\textit{\textbf{p}}_{t+\Delta t}) \coloneqq (\textit{\textbf{u}}(\textit{\textbf{q}}_{t},\textit{\textbf{p}}_{t}),\textit{\textbf{v}}(\textit{\textbf{q}}_{t},\textit{\textbf{p}}_{t})). 
\end{eqnarray}
Following the transformations $\mathbf{Q}(\textit{\textbf{q}},\textit{\textbf{p}})$ and $\mathbf{P}(\textit{\textbf{q}},\textit{\textbf{p}})$ in Eq. $\eqref{transeq}$, these equations can be rewritten as
\begin{eqnarray}
    {\textit{\textbf{Q}}}_{T+\Delta T} &=& {\textit{\textbf{Q}}}(\textit{\textbf{q}}_{t+\Delta t},\textit{\textbf{p}}_{t+\Delta t})\nonumber\\
    \label{canf1}
    &=& \textit{\textbf{u}'}(\textit{\textbf{Q}}_{T},\textit{\textbf{P}}_{T}) \coloneqq {\textit{\textbf{Q}}}\left[\textit{\textbf{u}}\left(\textit{\textbf{q}}(\textit{\textbf{Q}}_{T},\textit{\textbf{P}}_{T}),\textit{\textbf{p}}(\textit{\textbf{Q}}_{T},\textit{\textbf{P}}_{T})\right),\textit{\textbf{v}}\left(\textit{\textbf{q}}(\textit{\textbf{Q}}_{T},\textit{\textbf{P}}_{T}),\textit{\textbf{p}}(\textit{\textbf{Q}}_{T},\textit{\textbf{P}}_{T})\right)\right],\\
    {\textit{\textbf{P}}}_{T+\Delta T} &=& {\textit{\textbf{P}}}(\textit{\textbf{q}}_{t+\Delta t},\textit{\textbf{p}}_{t+\Delta t})\nonumber\\
    \label{canf2}
    &=& \textit{\textbf{v}'}(\textit{\textbf{Q}}_{T},\textit{\textbf{P}}_{T}) \coloneqq 
    {\textit{\textbf{P}}}\left[\textit{\textbf{u}}\left(\textit{\textbf{q}}(\textit{\textbf{Q}}_{T},\textit{\textbf{P}}_{T}),\textit{\textbf{p}}(\textit{\textbf{Q}}_{T},\textit{\textbf{P}}_{T})\right),\textit{\textbf{v}}\left(\textit{\textbf{q}}(\textit{\textbf{Q}}_{T},\textit{\textbf{P}}_{T}),\textit{\textbf{p}}(\textit{\textbf{Q}}_{T},\textit{\textbf{P}}_{T})\right)\right],
\end{eqnarray}
where $T = Q_0$, $\Delta T = \Delta Q_0$. % and 
%\begin{eqnarray}
%\textit{\textbf{Q}}_{T} - \textit{\textbf{q}}_{t} = (\textit{\textbf{Q}}_{T} - \textit{\textbf{Q}}_{t}) + (\textit{\textbf{Q}}_{t} - \textit{\textbf{q}}_{t}),\\
%\textit{\textbf{P}}_{T} - \textit{\textbf{p}}_{t} = (\textit{\textbf{P}}_{T} - \textit{\textbf{P}}_{t}) + (\textit{\textbf{P}}_{t} - \textit{\textbf{p}}_{t}).
%\end{eqnarray}
For the transformation $(\textit{\textbf{Q}},\textit{\textbf{P}}) = \left(\textit{\textbf{Q}}(\textit{\textbf{q}},\textit{\textbf{p}}),\textit{\textbf{P}}(\textit{\textbf{q}},\textit{\textbf{p}})\right)$ to be a canonical transformation, the following conditions must be satisfied:
\begin{eqnarray}
\label{condcond1}
 \textit{\textbf{u}'}(\textit{\textbf{Q}}_{T},\textit{\textbf{P}}_{T}) &\equiv&  \frac{\partial{H'(\textit{\textbf{Q}}_{T},\textit{\textbf{P}}_{T})}}{\partial{\textit{\textbf{P}}}_{T}}\Delta T + {\textit{\textbf{Q}}}_{T},\\
\label{condcond2}
 \textit{\textbf{v}'}(\textit{\textbf{Q}}_{T},\textit{\textbf{P}}_{T}) &\equiv& 
 -\frac{\partial{H'(\textit{\textbf{Q}}_{T},\textit{\textbf{P}}_{T})}}{\partial{\textit{\textbf{Q}}}_{T}}\Delta T + {\textit{\textbf{P}}}_{T}.
\end{eqnarray}
If $H$ and $H'$ are identically equal, the conditions of Eqs.~\eqref{condcond1} and \eqref{condcond2} are equivalent to 
\begin{equation}
\begin{split}
\label{proof_motion1}
 \textit{\textbf{u}'}(\textit{\textbf{q}}_{t},\textit{\textbf{p}}_{t}) \equiv  \textit{\textbf{u}}(\textit{\textbf{q}}_{t},\textit{\textbf{p}}_{t}),\\
 \textit{\textbf{v}'}(\textit{\textbf{q}}_{t},\textit{\textbf{p}}_{t}) \equiv 
 \textit{\textbf{v}}(\textit{\textbf{q}}_{t},\textit{\textbf{p}}_{t}).
\end{split}
\end{equation}
Eq.~\eqref{proof_motion1} is equivalent to the following condition (see Appendix~\ref{appendix_b}):
\begin{eqnarray}
&\:&
\left\{\textit{\textbf{q}}_{t+\Delta t},\textit{\textbf{p}}_{t+\Delta t},\textit{\textbf{q}}_{t},\textit{\textbf{p}}_{t}\:\mid \: (\textit{\textbf{q}}_{t+\Delta t},\textit{\textbf{p}}_{t+\Delta t}) = \left(\textit{\textbf{u}}(\textit{\textbf{q}}_{t},\textit{\textbf{p}}_{t}),\textit{\textbf{v}}(\textit{\textbf{q}}_{t},\textit{\textbf{p}}_{t})\right)\right\}\nonumber\\
&\:&\:= \left\{\textit{\textbf{Q}}_{T+\Delta T},\textit{\textbf{P}}_{T+\Delta T},\textit{\textbf{Q}}_{T},\textit{\textbf{P}}_{T}\:\mid \: (\textit{\textbf{q}}_{t+\Delta t},\textit{\textbf{p}}_{t+\Delta t}) = \left(\textit{\textbf{u}}(\textit{\textbf{q}}_{t},\textit{\textbf{p}}_{t}),\textit{\textbf{v}}(\textit{\textbf{q}}_{t},\textit{\textbf{p}}_{t})\right)\right\}.
\label{proofed_target2}
\end{eqnarray}
The time-series dataset $D$ is understood as the part of the subspace given on the left side of Eq.~\eqref{proofed_target2}.
\par
%By discretizing the time, the transformation that does not change the equation of motion satisfies the following conditions: 
%$\forall{(\textit{\textbf{q}}(t),\textit{\textbf{p}}(t))}$, $\{\textit{\textbf{q}}(t+\Delta t),\textit{\textbf{p}}(t+\Delta t)\:\mid \: \frac{\partial{H(\textit{\textbf{q}}(t),\textit{\textbf{p}})}(t)}{\partial{\textit{\textbf{q}}}(t)} = -\left( {\textit{\textbf{p}}}(t+\Delta t) - {\textit{\textbf{p}}}(t)\right), \:\frac{\partial{H(\textit{\textbf{q}}(t),\textit{\textbf{p}}(t))}}{\partial{\textit{\textbf{p}}(t)}} = \textit{\textbf{q}}(t+\Delta t) - \textit{\textbf{q}}(t)\}= \{\textit{\textbf{Q}}(t+\Delta t),\textit{\textbf{P}}(t+\Delta t)\:\mid \: \frac{\partial{H(\textit{\textbf{q}}(t),\textit{\textbf{p}})}(t)}{\partial{\textit{\textbf{q}}}(t)} = -\left( {\textit{\textbf{p}}}(t+\Delta t) - {\textit{\textbf{p}}}(t)\right), \:\frac{\partial{H(\textit{\textbf{q}}(t),\textit{\textbf{p}}(t))}}{\partial{\textit{\textbf{p}}(t)}} = \textit{\textbf{q}}(t+\Delta t) - \textit{\textbf{q}}(t)\}$. 
%従って，運動方程式とハミルトニアンを不変とする変換は，次の条件を満たすことと等価である．

\subsection{Noether's theorem and time-series dataset}
\label{noether_time}
By combining the conditions obtained in the previous two subsections, we obtain the condition that the Hamiltonian and canonical equations are simultaneously invariant under the transformation. The condition is acquired as
\begin{equation}
\begin{split}
\label{cond1}
 &\forall{E},\: \left\{\textit{\textbf{q}}_{t+\Delta t},\textit{\textbf{p}}_{t+\Delta t},\textit{\textbf{q}}_t,\textit{\textbf{p}}_t\:\middle|\: H(\textit{\textbf{q}}_t,\textit{\textbf{p}}_t) = E,  {\textit{\textbf{p}}}_{t+\Delta t} = {\textit{\textbf{p}}}_t-\frac{\partial{H(\textit{\textbf{q}}_t,\textit{\textbf{p}}_t)}}{\partial{\textit{\textbf{q}}}_t}, \textit{\textbf{q}}_{t+\Delta t} = \textit{\textbf{q}}_t + \frac{\partial{H(\textit{\textbf{q}}_t,\textit{\textbf{p}}_t)}}{\partial{\textit{\textbf{p}}_t}}\right\}\\
 &= \left\{\textit{\textbf{Q}}_{T+\Delta T},\textit{\textbf{P}}_{T+\Delta T},\textit{\textbf{Q}}_{T},\textit{\textbf{P}}_{T}\:\middle|\: H(\textit{\textbf{q}}_t,\textit{\textbf{p}}_t) = E,  {\textit{\textbf{p}}}_{t+\Delta t} = {\textit{\textbf{p}}}_t-\frac{\partial{H(\textit{\textbf{q}}_t,\textit{\textbf{p}}_t)}}{\partial{\textit{\textbf{q}}}_t}, \textit{\textbf{q}}_{t+\Delta t} = \textit{\textbf{q}}_t + \frac{\partial{H(\textit{\textbf{q}}_t,\textit{\textbf{p}}_t)}}{\partial{\textit{\textbf{p}}_t}}\right\}.
\end{split}
\end{equation}
If the time-series dataset $D$ has all possible data points under the Hamiltonian $H(\textit{\textbf{q}},\textit{\textbf{p}})$ and the canonical equations, $D$ is equivalent to the subspace defined on the left side of Eq.~\eqref{cond1}. 
Thus, the symmetry of the Hamilton system is associated with the symmetry of the time series dataset $D$. 
The transformation set satisfying Eq.~\eqref{cond1}, $\left\{\textit{\textbf{Q}}(\textit{\textbf{q}},\textit{\textbf{p}}),\textit{\textbf{P}}(\textit{\textbf{q}},\textit{\textbf{p}})\:\middle|\: {\rm satisfy}\:\:{\rm Eq.}\: \eqref{cond1}\right\}$, is the same as the invariant transformation set  $\mathbbm{c}_{\rm inv}:\:\left\{\mathbf{\mathcal{Q}}(\textit{\textbf{q}},\textit{\textbf{p}}, \boldsymbol{\theta}),\mathbf{\mathcal{P}}(\textit{\textbf{q}},\textit{\textbf{p}}, \boldsymbol{\theta})\:\middle|\: \boldsymbol{\theta} \in \mathbb{R}^{d_{\theta}} \right\}$ under the discretized equations of motion.\par 
The transformed dataset in Eq.~\eqref{cond1},
\begin{equation}
\left\{\textit{\textbf{Q}}_{T+\Delta T},\textit{\textbf{P}}_{T+\Delta T},\textit{\textbf{Q}}_{T},\textit{\textbf{P}}_{T}\:\middle|\: H(\textit{\textbf{q}}_t,\textit{\textbf{p}}_t) = E,  {\textit{\textbf{p}}}_{t+\Delta t} = {\textit{\textbf{p}}}_t-\frac{\partial{H(\textit{\textbf{q}}_t,\textit{\textbf{p}}_t)}}{\partial{\textit{\textbf{q}}}_t}, \textit{\textbf{q}}_{t+\Delta t} = \textit{\textbf{q}}_t + \frac{\partial{H(\textit{\textbf{q}}_t,\textit{\textbf{p}}_t)}}{\partial{\textit{\textbf{p}}_t}}\right\},
\end{equation}
is obtained by the time evolution $t \to T$ of time-series dataset at $t$:
\begin{equation}
\left\{\textit{\textbf{Q}}_{t+\Delta t},\textit{\textbf{P}}_{t+\Delta t},\textit{\textbf{Q}}_{t},\textit{\textbf{P}}_{t}\:\middle|\: H(\textit{\textbf{q}}_t,\textit{\textbf{p}}_t) = E,  {\textit{\textbf{p}}}_{t+\Delta t} = {\textit{\textbf{p}}}_t-\frac{\partial{H(\textit{\textbf{q}}_t,\textit{\textbf{p}}_t)}}{\partial{\textit{\textbf{q}}}_t}, \textit{\textbf{q}}_{t+\Delta t} = \textit{\textbf{q}}_t + \frac{\partial{H(\textit{\textbf{q}}_t,\textit{\textbf{p}}_t)}}{\partial{\textit{\textbf{p}}_t}}\right\}. 
\end{equation}
If the Hamiltonian is given, we can obtain the time-evolved dataset by evolving the dataset obeying the canonical equations of motion. Even if the Hamiltonian is not given, we can obtain a time-evolved dataset as follows. Assume that we have time-series dataset at
$(t, t + \Delta t, t + 2\Delta t, \dots , t + s\Delta t,\dots )$, where $s$ is $\mathbb{Z}_{\geq 0}$. The time transformation of data from $t$ to $T$ can be approximated by replacing $T$ with $T'$: 
\begin{eqnarray}
T' &=& t + s\Delta t,\\
s &=& \argmin_{s} \left|T-(t + s\Delta t)\right|.
\end{eqnarray}\par
There is no guarantee that all energy states in the reduced Hamiltonian are realized in the original complex system. 
In particular, when constructing a reduced model of a metastable state, only its energy state is realized. 
%Collective motionの縮約モデルにおいて，そのハミルトニアンで定義される全てのエネルギー状態が実現される保証はない．また，準安定状態の縮約モデルを構築する場合には，そもそもそのエネルギー状態以外はとることはない．
%The condition Eq.~\eqref{cond1} can be re-expressed as a union of the divided conditions as explained below. 
%これは，
%$\textit{\textbf{q}}(t+\Delta t),\textit{\textbf{p}}(t+\Delta t),\textit{\textbf{q}}(t),\textit{\textbf{p}}(t)$に力学系が取り得る全ての状態が作る部分空間の座標変換に対する不変性が，ハミルトニアンと運動方程式を不変とする変換であることを意味する．
To overcome this difficulty, we introduce the different expressions of the condition in Eq.~\eqref{cond1}. 
Let $E_i$ be a real number representing one energy state. 
%が成り立つような変換をA_i(\theta)とおくと，
%全てのEで成り立つ変換は
%UA_iとなる．
We also define the transformation
\begin{eqnarray}
\mathbbm{c}_i:\: \Gamma \times \mathbb{R} \times \mathbb{R} &\longrightarrow& \Gamma \times \mathbb{R} \times \mathbb{R},\\ 
\:(\textit{\textbf{q}},\textit{\textbf{p}}) &\longmapsto& (\mathbf{Q},\mathbf{P}) \coloneqq \textbf{(}\mathbf{Q}_i(\textit{\textbf{q}},\textit{\textbf{p}}),\mathbf{P}_i(\textit{\textbf{q}},\textit{\textbf{p}})\textbf{)},
\end{eqnarray}
which satisfy 
\begin{equation}
\begin{split}
&\left\{\textit{\textbf{q}}_{t+\Delta t},\textit{\textbf{p}}_{t+\Delta t},\textit{\textbf{q}}_t,\textit{\textbf{p}}_t\:\middle|\: H(\textit{\textbf{q}}_t,\textit{\textbf{p}}_t) = E_i,  {\textit{\textbf{p}}}_{t+\Delta t} = {\textit{\textbf{p}}}_t-\frac{\partial{H(\textit{\textbf{q}}_t,\textit{\textbf{p}}_t)}}{\partial{\textit{\textbf{q}}}_t}, \textit{\textbf{q}}_{t+\Delta t} = \textit{\textbf{q}}_t + \frac{\partial{H(\textit{\textbf{q}}_t,\textit{\textbf{p}}_t)}}{\partial{\textit{\textbf{p}}_t}}\right\}\\
 &= \left\{\textit{\textbf{Q}}_{T+\Delta T},\textit{\textbf{P}}_{T+\Delta T},\textit{\textbf{Q}}_{T},\textit{\textbf{P}}_{T}\:\middle|\: H(\textit{\textbf{q}}_t,\textit{\textbf{p}}_t) = E_i,  {\textit{\textbf{p}}}_{t+\Delta t} = {\textit{\textbf{p}}}_t-\frac{\partial{H(\textit{\textbf{q}}_t,\textit{\textbf{p}}_t)}}{\partial{\textit{\textbf{q}}}_t}, \textit{\textbf{q}}_{t+\Delta t} = \textit{\textbf{q}}_t + \frac{\partial{H(\textit{\textbf{q}}_t,\textit{\textbf{p}}_t)}}{\partial{\textit{\textbf{p}}_t}}\right\}.
 \label{ei_cond1}
\end{split}
\end{equation}
Because the invariant transformation that satisfies Eq.~\eqref{eq_hamiltonian_condition} does not change the energy, the condition Eq.~\eqref{cond1} can be re-expressed as a union of the divided conditions:  $\left\{\textit{\textbf{Q}}(\textit{\textbf{q}},\textit{\textbf{p}}),\textit{\textbf{P}}(\textit{\textbf{q}},\textit{\textbf{p}})\:\middle|\: {\rm satisfy}\:\:{\rm Eq.}\: \eqref{cond1}\right\}=\bigcap_{i}\left\{\textit{\textbf{Q}}_i(\textit{\textbf{q}},\textit{\textbf{p}}),\textit{\textbf{P}}_i(\textit{\textbf{q}},\textit{\textbf{p}})\:\middle|\: {\rm satisfy}\:\:{\rm Eq.}\: \eqref{ei_cond1}\right\}$. 
%これは，あるエネルギーE_iについての成り立つ不変性が，
%系の不変性の候補になることを示す．
This implies that the invariant transformation set for a certain energy $E_i$ must include some invariant transformations for the total energy. 
Thus, candidate transformations that make the Hamiltonian and canonical equations invariant are obtained as the transformations that make the subspace
\begin{eqnarray}
S_i\coloneqq\left\{\textit{\textbf{q}}_{t+\Delta t},\textit{\textbf{p}}_{t+\Delta t},\textit{\textbf{q}}_t,\textit{\textbf{p}}_t\:\middle|\: H(\textit{\textbf{q}}_t,\textit{\textbf{p}}_t) = E_i,  {\textit{\textbf{p}}}_{t+\Delta t} = {\textit{\textbf{p}}}_t-\frac{\partial{H(\textit{\textbf{q}}_t,\textit{\textbf{p}}_t)}}{\partial{\textit{\textbf{q}}}_t}, \textit{\textbf{q}}_{t+\Delta t} = \textit{\textbf{q}}_t + \frac{\partial{H(\textit{\textbf{q}}_t,\textit{\textbf{p}}_t)}}{\partial{\textit{\textbf{p}}_t}}\right\}
\label{noether_manifold}
\end{eqnarray}
invariant. 
This expression is useful to find the candidates of symmetries in a complex dynamical system, such as dynamics at the metastable state.\par
In a finite time measurement or simulation, only data $D$ of a subset of $S_i$ can be obtained. 
On the basis of the following two physical principles, we can estimate $S_i$ from data $D$. The first principle is described as follows. 
The subspace $S_i$ can be represented as a product space of two subspaces:
\begin{eqnarray}
S_i &=& S_i^a \times S_i^b,\\
S_i^a &=& \left\{\textit{\textbf{q}}_t,\textit{\textbf{p}}_t\:\middle|\: H(\textit{\textbf{q}}_t,\textit{\textbf{p}}_t) = E_i,  {\textit{\textbf{p}}}_{t+\Delta t} = {\textit{\textbf{p}}}_t-\frac{\partial{H(\textit{\textbf{q}}_t,\textit{\textbf{p}}_t)}}{\partial{\textit{\textbf{q}}}_t}, \textit{\textbf{q}}_{t+\Delta t} = \textit{\textbf{q}}_t + \frac{\partial{H(\textit{\textbf{q}}_t,\textit{\textbf{p}}_t)}}{\partial{\textit{\textbf{p}}_t}}\right\}\\ &=& \left\{\textit{\textbf{q}}_t,\textit{\textbf{p}}_t\:\middle|\: H(\textit{\textbf{q}}_t,\textit{\textbf{p}}_t) = E_i\right\},\\
S_i^b &=&\left\{\textit{\textbf{q}}_{t+\Delta t},\textit{\textbf{p}}_{t+\Delta t}\:\middle|\: H(\textit{\textbf{q}}_t,\textit{\textbf{p}}_t) = E_i,  {\textit{\textbf{p}}}_{t+\Delta t} = {\textit{\textbf{p}}}_t-\frac{\partial{H(\textit{\textbf{q}}_t,\textit{\textbf{p}}_t)}}{\partial{\textit{\textbf{q}}}_t}, \textit{\textbf{q}}_{t+\Delta t} = \textit{\textbf{q}}_t + \frac{\partial{H(\textit{\textbf{q}}_t,\textit{\textbf{p}}_t)}}{\partial{\textit{\textbf{p}}_t}}\right\}.\\
\end{eqnarray}
%ハミルトニアンがC^2 classであることから，Aは微分可能多様体となる．
Since the Hamiltonian is a $C^2$ class function, $S_i^a$ is a differentiable manifold. 
%また，ハミルトニアンがC^2級であることから，正準運動方程式はC^1級となる．
The canonical equation of motion is a $C^1$ map because the Hamiltonian is a $C^2$ class function. 
%これより，Aの写像であるBも微分可能多様体となる．
The subspace $S_i^b$ is a subspace mapped from manifold $S_i^a$ according to the canonical equations of motion. 
Therefore, the subspace $S_i^b$ is also a differentiable manifold, and $S_i$ is the product of differentiable manifolds $S_i^a$ and $S_i^b$. 
From a property of product manifold, $S_i$ is understood as a differentiable manifold. 
%この事実に基づき，本研究では，深層学習のような機械学習によって，有限のデータからデータ補完を通して，多様体Sを近似的に抽出することを提案する．
Interpolation of differentiable manifolds can be realized by machine learning methods such as deep learning. 
In our proposed framework, $S_i$ is estimated from a finite number of data $D$ using a deep learning technique. 
%あるエネルギーに関する多様体S_iに関するデータを抽出したい場合，エネルギーが変化するような正準力学系では，多数の試行を必要とすると考えられる．しかし，
The second principle is described as follows. 
In a canonical dynamical system in which the energy changes with time, it is not efficient to acquire the data of $S_i$ because $S_i$ is a subspace of specific energy. 
The important cases of a complex dynamical system to be modeled as a reduced model are at the stable or metastable state. 
Also, one of the final goals of this study is to extract the conservation laws in a large-scale collective motion system at a metastable state. 
%このようなシステムでの運動ではエネルギーは保存されるので，
In the stable or metastable state, the energy of the system is conserved:  $H(\textit{\textbf{q}}_{t},\textit{\textbf{p}}_{t}) = H(\textit{\textbf{q}}_{t+\Delta t},\textit{\textbf{p}}_{t+\Delta t}) = E$. 
Therefore, for the purpose of this study, efficient data acquisition is realized.  
%従って，S_iを推定するためには，E_iを初期値とするシミュレーションを一度実行すれば，そこから複数の初期条件でのデータ点を獲得できると考えられる．
%また，このような系では，自明な時間発展の対称性を持つ．
%On the other hand, the purpose of this paper is to show that the proposed framework for estimating the conservation law is feasible, so we set the transformation of time as the identity mapping $t \to t$ for simplification. 
\par
%本研究では，シミュレーションや計測を通してこの部分空間の一部のデータ${q_n(t+\Delta t),p_n(t+\Delta t),q_n(t),p_n(t)}_{n=1}^{n=N}$が得られた場合に，DNNの多様体近似による補完によって近似的に部分空間を推定できることを仮定し，次節のデータ多様体の対称性抽出法によって，力学系の対称性を推定する．
%From observations or from computational simulations, let there be finite time series data $D$ which are a part of the subspace $S$. 
%From $D$, we assume that the subspace $S$ can be approximated by the DNN as a manifold, in addition to assuming that the invariant transformation for $S$ is estimated by the symmetry of the manifold. 
%この仮定によって得られた結果の妥当性は，
%最終的に得られた保存量が時間普遍であることを確認することで
%容易に確認される．
%This assumption can be easily violated when the number of data samples is not enough to reconstruct the $S$. Although, the conservation laws obtained based on this assumption can be easily verified by confirming whether the conserved value is invariant in the time-series data. \par
%%%%%%%%%%%%%%%%%%%%%%%%%%%%%%%%%%%%%%%%%%%%%%%%%%%%%
%<場の理論との関係>
%本研究では，古典系での保存則推定についてのみ扱った．
%しかし，本研究の枠組みは場の量子系へ適用することも可能である．
%場の量子論では，ハミルトニアンは，場$\phi$と座標$x = (ict,x_1,x_2,\cdots)$の関数
In this study, we only deal with classical systems. A similar relationship holds between the data manifold and the symmetry of the system in canonical quantum field theory. 
In the canonical quantum field theory, the Hamiltonian is given as
\begin{eqnarray}
H\left({\bm \phi}({\bm x}),{\bm \pi}({\bm x}),{\bm x}\right),
\end{eqnarray}
where ${\bm \phi}({\bm x})$ is the field, ${\bm \pi}({\bm x})$ is the canonical momentum conjugate of ${\bm \phi}({\bm x})$, and ${\bm x}=(ct, x_1, x_2, x_3)$ is the Minkowski space; ${\bm \phi}({\bm x})$ and ${\bm \pi}({\bm x})$ satisfy the commutation relation
\begin{eqnarray}
\left[{\bm \phi}({\bm x}), {\bm \pi}({\bm y})\right] &=& i\delta^{(4)}({\bm x} - {\bm y})\\
\left[{\bm \phi}({\bm x}), {\bm \phi}({\bm y})\right] &=& \left[{\bm \pi}({\bm x}), {\bm \pi}({\bm y})\right] = 0.
\end{eqnarray}

The infinitesimal transformation is given as
%となる．その上でそれらの変換
\begin{eqnarray}
\Phi^i({\bm X}) &=& \phi^i({\bm x}) + \delta \phi^i({\bm x}),\\
\Pi^i({\bm X}) &=& \pi^i({\bm x}) + \delta \pi^i({\bm x}),\\
X^i &=& x^i + \delta x^i. 
\end{eqnarray}
%古典系で座標と時刻が入れ子構造になっていたように，場の量子論の正準理論は場とその共役運動量は，ミンコフスキ空間を入れ子構造としてもる．
Similar to the nested relations between coordinates and time in the classical system, the canonical quantum field theory states that a field and its conjugate momentum have a nested Minkowski space. 
Therefore, as in the discussion for classical systems, 
the following relation is given as a condition of the invariant transformation of a Hamiltonian system: 
%この点に注意して，古典系と同様の議論を行うと，
%時系列データ多様体の対称性とハミルトニアンシステムの対称性の間に以下のような関係式が成り立つ．
\begin{eqnarray}
\forall E, &\:&\{ {\bm \phi}_{t+\Delta t},{\bm \pi}_{t+\Delta t},{\bm \phi}_t,{\bm \pi}_t\:\mid H({\bm \phi}_t,{\bm \pi}_t) = E,  ({\bm \phi}_{t+\Delta t}, {\bm \pi}_{t+\Delta t}) = \textit{\textbf{u}}({\bm \phi}_t,{\bm \pi}_t)\}\nonumber\\ 
&\:&= \{{\bm \Phi}_{T+\Delta T},{\bm \Pi}_{T+\Delta T},{\bm \Phi}_T,{\bm \Pi}_T\:\mid H({\bm \phi}_t,{\bm \pi}_t) = E,  ({\bm \phi}_{t+\Delta t}, {\bm \pi}_{t+\Delta t}) = \textit{\textbf{u}}({\bm \phi}_t,{\bm \pi}_t)\},\nonumber 
\end{eqnarray}
where $\textit{\textbf{u}}$ is an equation of motion such as the Klein–Gordon equation of a scalar particle. 

\subsection{DNN and data manifold}
\label{sec2}
%カオスのような状況を除いて，時系列データ集合（subspace S）は，多様体構造をなすと考えられる．
As mentioned in Sec.~\ref{noether_time}, the subspace $S_i$ could be modeled as a differentiable manifold using machine learning models. 
%A differentiable manifold is a space constructed by continuously pasting Euclidean spaces called tangent spaces. 
%An approximate example of a differentiable manifold is the earth. 
%The earth's surface is considered as a lamination of maps that are two-dimensional Euclidean spaces. 
Some well-trained DNNs have the ability to model the distribution of a training dataset as a differentiable manifold~\cite{Irie_1990,Hinton_Reducing_2006,brahma2016deep,basri2016efficient,Rifai_The_2011,mototakearob}. 
In this paper, we refer to such a differentiable manifold as a data manifold.\par
We explain how a DNN models a $d_{m}$-dimensional manifold in $d_{\rm in}$-dimensional space $\textit{\textbf{x}}$ using one of the simplest DNNs: a feed forward three-layer DNN, for which the input has $d_{\rm in}$ dimensions, the hidden layer has $d_{\rm h}(> d_{\rm in})$ dimensions, and the output has $d_{\rm out}(< d_{\rm in}) = d_m$ dimensions. The mapping function
$\textit{\textbf{f}}_{\rm DNN}(\textit{\textbf{x}})=\left[f_1(\textit{\textbf{x}}),f_2(\textit{\textbf{x}}),\cdots ,f_{d_{\rm out}}(\textit{\textbf{x}})\right]$ of the DNN is defined as $\textit{\textbf{f}}_{\rm DNN}(\textit{\textbf{x}}) =  \textit{\textbf{w}}^{h}\textit{\textbf{h}}
 = \textit{\textbf{w}}^{h}{\boldsymbol \varphi}(\textit{\textbf{w}}^{\rm in}\textit{\textbf{x}})
$, 
where $\textit{\textbf{h}}=(h_{1},h_{2},\cdots ,h_{d_{h}})$ is the $d_{h}$-dimensional output of the hidden layer. 
We define ${\boldsymbol \varphi}(\cdot )$ as
$
{\boldsymbol \varphi}(\textit{\textbf{w}}^{\rm in}\textit{\textbf{x}}) = (\varphi_1,\varphi_2, \cdots, \varphi_{d_{h}}), 
\varphi_j = \varphi\left[\sum_i^{d_{\rm in}} \left(w_{ij}^{\rm in}x_i\right)\right],
$
where $\varphi$ is the activation function. 
\begin{figure}
 \begin{center}
  \includegraphics[width=8.0cm]{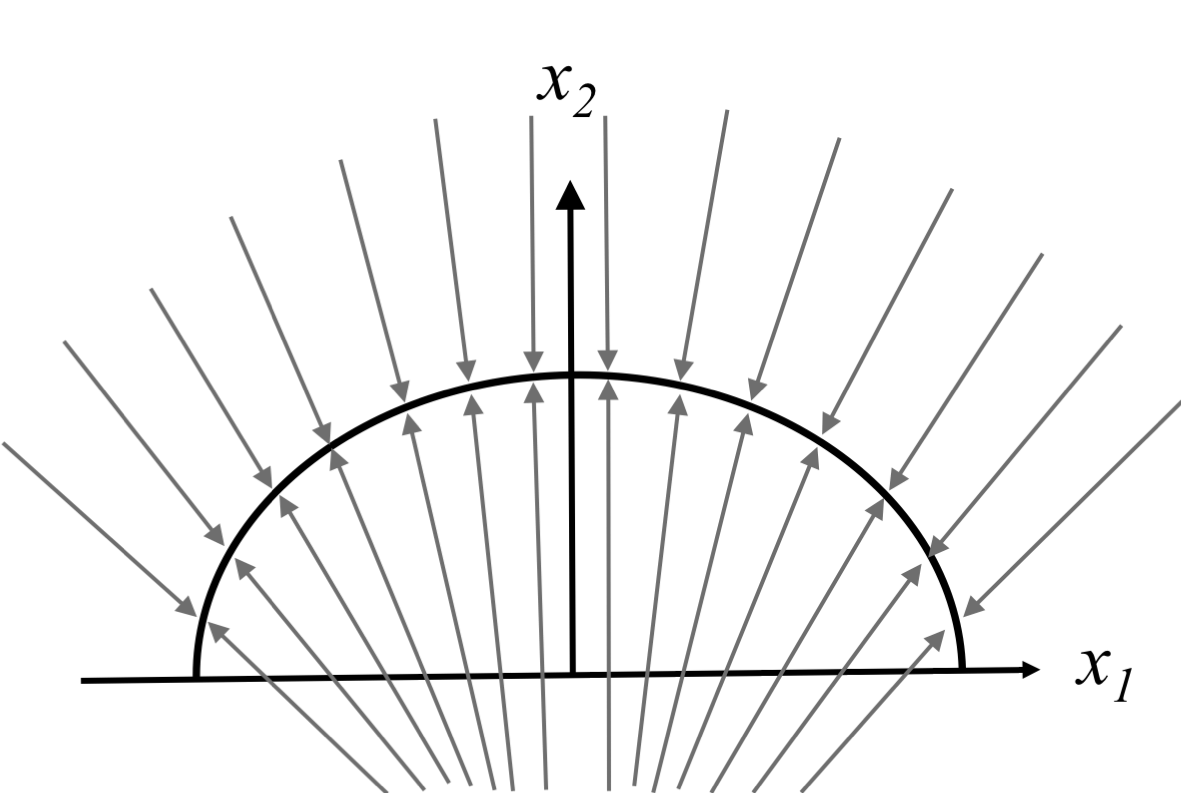}
  \caption{Schematic diagram of the mapping structure of a two-dimensional input space in a DNN trained with data distributed on a black curve. 
The arrows indicate the compression direction of the input space in the mapping from the input to the hidden layer.}
  \label{fig_1}
 \end{center}
\end{figure}
Usually, a sigmoid or ReLU function is used as the activation function. 
These activation functions are constructed using linear and flat domains. 
On the basis of these properties of activation functions, $\varphi_j$ maps the input subspace related to the linear domain of the activation function to a one-dimensional space to align the vector $(w_{0j},w_{1j}\cdots,w_{d_{\rm in}j})$. 
If the number of $\varphi_j$ sharing the same input subspace is $d_{\rm out}$, the $\varphi_j$ defines a $d_{\rm out}$-dimensional sub-hyperplane. 
The DNN models the data distribution by continuously pasting these sub-hyperplanes as if they were the tangent spaces of a data manifold. 
That is, the DNN embeds the input space in the output space by pasting the sub-hyperplanes and compresses the tangent direction of these sub-hyperplanes (Fig.~\ref{fig_1}). 
Deeper and more complex DNNs can be understood as a collection of such three-layer DNN. 
Thus, such deeper DNNs can model more complex manifold structures as a combination of simple manifold structures modeled by a three-layer DNN~\cite{basri2016efficient}. 
Note that the output of a three-layer DNN, a part of the deeper DNN, is referred to as a hidden layer. 
This is only one example of how a DNN models a data manifold. However, many studies have suggested that there are resemble property in successful trained DNNs~\cite{Irie_1990,Hinton_Reducing_2006,brahma2016deep,basri2016efficient,Rifai_The_2011,mototakearob}. By replacing the input space from $\textit{\textbf{x}}$ to $\Gamma \times \mathbb{R} \times \mathbb{R}$, we can also model a time-series data manifold $S_i$ using DNN.\par
%本研究では，データ多様体がDNNによってモデル化されたことを前提に，そこからシステムの対称性を抽出する手法を提案する．
In this study, using a trained DNN that models a time-series data manifold $S_i$, we propose a method of extracting information about the symmetry of a dynamical system. 
%DNN以外にも多様体をモデル化できる手法は存在するが，近年のDNNを用いた物理データ分析の莫大な知見を活用することを念頭にDNNを用いる．
As described later in Sec.~\ref{summary}, our proposed framework does not require special DNNs, so we can directly utilize the vast knowledge obtained from studies on physical data analysis using DNNs.
This is why we select the DNN from multiple machine learning models that can be used to model manifolds.

\section{Method}
\begin{figure}[htbp]
  \begin{center}
   \includegraphics[width=160mm]{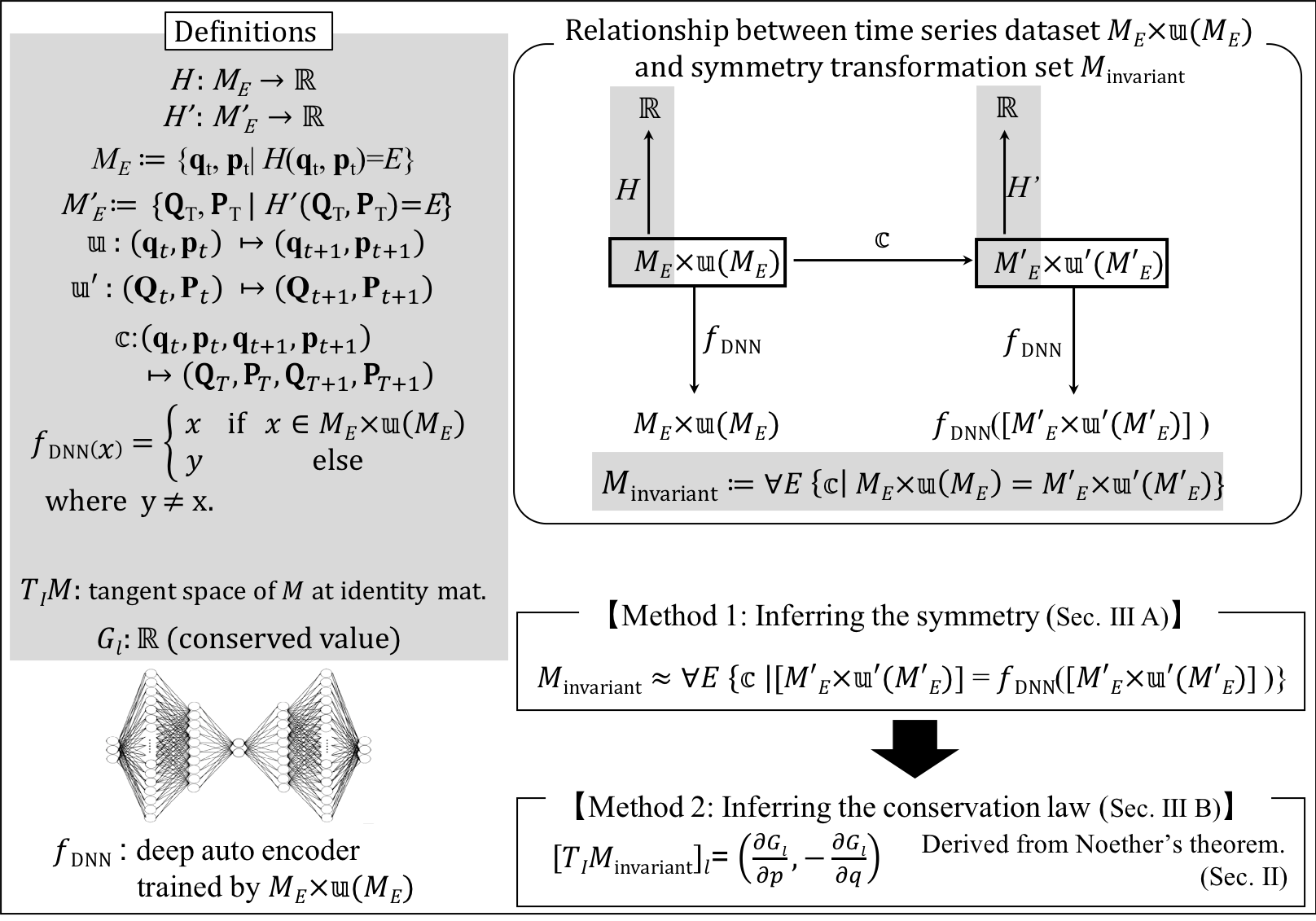}
  \caption{Schematic diagram of the proposed framework.}
  \label{fig_abst}
  \end{center}
\end{figure}

%本節では，時系列データから保存則を推定する手法の説明を行う．
In this section, we describe our proposed framework for estimating the conservation law from a time-series dataset of dynamics. 
%手法の概念図をFig.に示す．
The schematic diagram of the proposed framework is shown in Fig.~\ref{fig_abst}. 
The framework consists of two methods. 
%前節でのSec.~\ref{sec_noether} gives the derivation of the relationship between the symmetry of the time-series data distribution and the conservation law using Noether's theorem. に基づき，節では，　し　節ではをする．
In Sec.~\ref{sec_extract}, on the basis of the derivation of the relationship between the symmetry of the time-series dataset distribution and the conservation law (Sec.~\ref{theory}), we propose a method of inferring the symmetry of data manifold using the Monte Carlo sampling method. 
%In Sec.~\ref{sec_consest}, also describes another proposed method of inferring the conservation law from the obtained symmetry. 
In Sec.~\ref{sec_consest}, we describe the proposed method of inferring the conservation law from the obtained symmetry. 

\subsection{Method 1: Inferring the symmetry of data manifold using Monte Carlo sampling method}
\label{sec_extract}
\begin{figure}
 \begin{center}
    \includegraphics[clip,width=14.0cm]{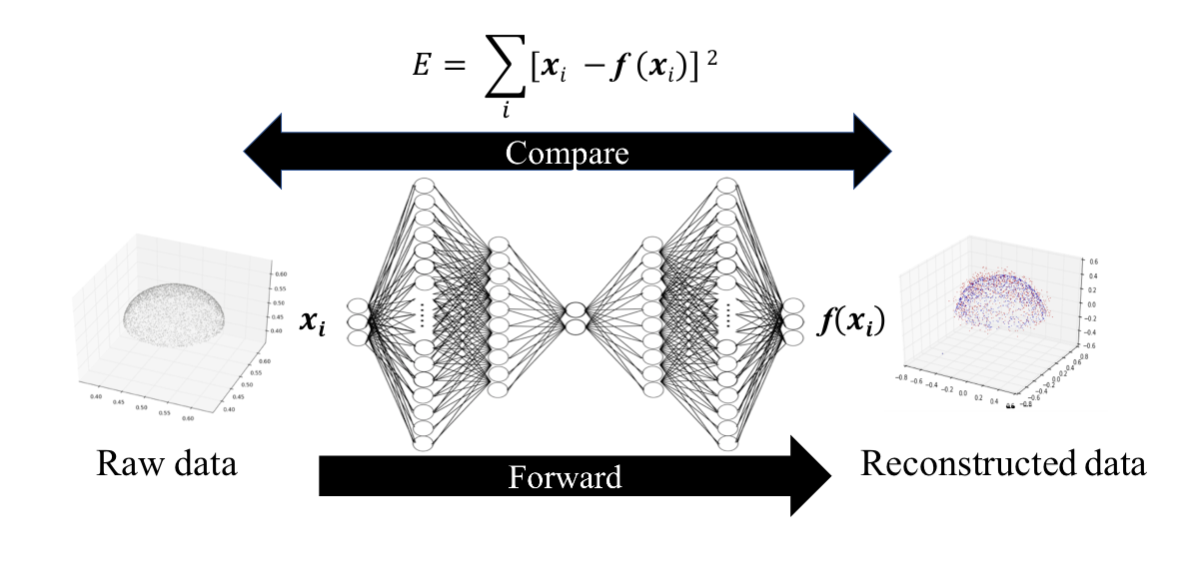}
    \caption{Schematic diagram of method of extracting invariant transformation using autoencoder.}
    \label{fig_2}
 \end{center}
\end{figure}
%本節では，物理系の時系列データに限定されない，データ多様体が持つ対称性を抽出する一般的な手法を提案する．
In this subsection, we propose a general method of inferring the symmetric property of data manifolds, which is not limited to the physical time-series dataset. 
%前節の議論より，
%入力空間で多様体上にないデータ点は，
%多様体上に吸引されるとわかる．
It can be inferred from the discussion in Sec.~\ref{sec2} that data points that are not on the manifold in the input space are attracted to the manifold (Fig.~\ref{fig_1}). 
%中間層において一度多様体上に吸引されたデータ点は，
%それよりも後の層において多様体上のデータと分離される
%ことはない．
Once the data points are attracted to the manifold in the hidden layer, they continue to exist on the manifold in the output $\textit{\textbf{f}}({\bf x})$. 
%したがって，出力層$\textit{\textbf{F}}({\bf x})$において，全てのデータは多様体上，あるいは空間の角
%となる0と1の組み合わせで与えられる
%座標点に縮退する．
%Therefore, in the output layer $\textit{\textbf{F}}({\bf x})$, the distribution of data embedded on the manifold or the corners of the space.\par
We propose a method based on this property of DNNs for extracting the symmetry of the data manifold using a deep autoencoder~\cite{Hinton_Reducing_2006}. 
The deep autoencoder is a model that compresses the input space to a low-dimensional hidden layer and decompresses the layer to an output space with the same dimension as the input space. 
In the decompression process, only the subspace of the input space around the data manifold is recovered because of the DNN property. 
%これをふまえると，
%あるデータ多様体を学習したDNNに対して，
%新たに与えられたデータセット$\{\textit{\textbf{x}}_i\}_{i=1}^N$
%がそのデータ多様体上にあるかは，
%それが出力空間に写像された際のデータ集合$\{\textit{\textbf{F}}(\textit{\textbf{x}}_i)\}_{i=1}^N$
%との間の二乗誤差
%\begin{equation}
%E(\{\textit{\textbf{x}}_i\}_{i=1}^N;\textit{\textbf{F}}(\textit{\textbf{x}})) = \sum_{i=1}^N %\left[ \textit{\textbf{x}}_i - \textit{\textbf{F}}(\textit{\textbf{x}}_i)\right]^2
%\end{equation}
%によって評価されるとわかる（図\ref{fig_innerdnn}）．
On the basis of this property, we can evaluate whether a transformation $\textit{\textbf{X}}(\cdot)$ causes the dataset distribution $\{ \textit{\textbf{x}}_i\}_{i=1}^N$ to remain in the same subspace of the data manifold (Fig.~\ref{fig_2}). 
The procedure is as follows. 
First, we train the deep autoencoder using $\{ \textit{\textbf{x}}_i\}_{i=1}^N$ as a training dataset. 
Second, we input the transformed dataset $\left\{\textit{\textbf{X}}( \textit{\textbf{x}}_i)\right\}_{i=1}^N$ into the trained deep autoencoder. Note that the deep autoencoder is not trained on the transformed dataset. 
Third, we evaluate the transformation $\textit{\textbf{X}}(\cdot)$ using the mean squared error between the input distribution of the dataset and its mapped distribution: 
\begin{equation}
\label{eqdnnm}
E_{\rm samp}[\textit{\textbf{X}}(\cdot)] = \frac{1}{N}\sum_{i=1}^N \left\{ \textit{\textbf{X}}(\textit{\textbf{x}}_i) - \textit{\textbf{f}}_{\rm DNN}[\textit{\textbf{X}}(\textit{\textbf{x}}_i)]\right\}^2. 
\end{equation}
A smaller $E_{\rm samp}$ value implies that $\textit{\textbf{X}}(\cdot)$ is a more invariant transformation. 
%DNNの訓練時に用いたデータ多様体分布が
%どのような座標変換${\bf A}$に対して普遍となるか
%ということである．
Using the criterion $E_{\rm samp}$, we approximate the invariant transformation set as
\begin{eqnarray}
\left\{\textit{\textbf{X}}(\cdot) \middle|\: \argmin_{\textit{\textbf{X}}}  E_{\rm samp}[\textit{\textbf{X}}(\cdot)]\right\}.
\end{eqnarray}\par
To infer the conservation law, it is necessary to estimate the invariant transformation set $M_{{\rm invariant}}$ of the manifold $S_i$. 
The invariant transformation set $M_{{\rm invariant}}$ is defined as 
\begin{equation}
M_{{\rm invariant}} \coloneqq \left\{\mathbf{\mathcal{Q}}^{S_i}(\cdot,\cdot, \boldsymbol{\theta}),\mathbf{\mathcal{P}}^{S_i}(\cdot,\cdot, \boldsymbol{\theta}) \middle|\:\boldsymbol{\theta}\right\}. 
\end{equation}
In Eq.~\eqref{eqdnnm}, by substituting $ \{\textit{\textbf{x}}_i\}_{i=1}^N$ for  $D = \left\{\textit{\textbf{q}}_{t_i}^i, \textit{\textbf{p}}_{t_i}^i,\textit{\textbf{q}}_{t_i+\Delta t}^i, \textit{\textbf{p}}_{t_i+\Delta t}^i\right\}_{i=1}^N$ and $\textit{\textbf{X}}(\cdot)$ for the transformation $\mathbbm{c}: \left(\textit{\textbf{Q}}(\cdot,\cdot),\textit{\textbf{P}}(\cdot,\cdot)\right)$, 
we can approximate $M_{{\rm invariant}}$ as
\begin{eqnarray}
M_{{\rm invariant}} \sim   \left\{\textit{\textbf{Q}}(\cdot,\cdot),\textit{\textbf{P}}(\cdot,\cdot)\middle|\:\argmin_{\textit{\textbf{Q}}(\cdot,\cdot),\textit{\textbf{P}}(\cdot,\cdot)}  E_{\rm samp}\left[\textit{\textbf{Q}}(\cdot,\cdot),\textit{\textbf{P}}(\cdot,\cdot)\right]\right\},
\end{eqnarray}
where dataset $D$ is generated from dynamics data at energy $E_i$. 
The approximated invariant transformation set is obtained approximately by sampling from the probabilistic density:
\begin{eqnarray}
{\rm P}\left(\textit{\textbf{Q}}(\cdot,\cdot),\textit{\textbf{P}}(\cdot,\cdot)\right) \sim \frac{1}{Z}\exp\left\{-\frac{N}{2\sigma^2}E_{\rm samp}\left[\textit{\textbf{Q}}(\cdot,\cdot),\textit{\textbf{P}}(\cdot,\cdot)\right]\right\},
\end{eqnarray}
where $\sigma$ is set as small as necessary and $Z$ is a normalization constant. 
Note that to actually perform this sampling, it is necessary to first give a concrete coordinate system of $(\textit{\textbf{q}}_i, \textit{\textbf{p}}_i)$ in which physicists want to search conservation laws.\par
%On the other hand, the transformation $\textit{\textbf{X}}(\cdot)$ treated in this method is not limited in a continuous transformation. 
As mentioned in Sec.~\ref{sec_noether}, continuous symmetries form a Lie group. 
Using the continuous parameter set $\boldsymbol{\theta} = \{\theta_k \}_{k=1}^{d_{\theta}}$, we define the representation of the Lie group as a $2d\times 2d$-dimensional matrix 
%リー郡の表現は，滑らかなパラメータ集合$\boldsymbol{\theta} = \{\theta_k\}_{k=1}^p$を用いて，
$A_{ij}(\boldsymbol{\theta}) = a_{ij}(\boldsymbol{\theta})$, where $2d$ is the degree of freedom of the target Hamiltonian system. 
$\boldsymbol{\theta}$ is a continuous parameter set, and $A(\boldsymbol{0}) = \boldsymbol{I}$. 
%したがって，Lie群の表現に限定すれば，
%行列Aを推定することが問題となる．
In the following, candidate invariant transformations are searched for within the Lie group representations. 
%For this purpose, we set the linear transformation $A = a_{jk}$ as a candidate for the invariant transformation. 
%具体的には，行列${\bf A}$の$d_{\rm in}\times d_{\rm in}$個ある要素$a_{jk}$についてサンプリングを行うことで，
%$E(\{{\bf A}\textit{\textbf{x}}_i\}_{i=1}^N;\textit{\textbf{F}}({\bf A}\textit{\textbf{x}}))$がある閾値以下となる${\bf A}$
%の集合が得られる．
The invariant transformation is obtained by sampling an element $a_{jk}$ of the matrix $A$ following the probability distribution
\begin{equation}
\label{samp_func0}
{\rm P}(a_{11},a_{12},a_{21},\cdots,a_{2d\:2d}) = \frac{1}{Z} \exp\left[-\frac{N}{2\sigma^2} E_{\rm samp}(a_{11},a_{12},a_{21},\cdots,a_{2d\:2d})\right]. 
\end{equation}\par
%このサンプリングを実行するには，$\sigma$を指定する必要があるが，
%事前に$\sigma$はわからない．
To perform this sampling, we need to specify $\sigma$. Ideally, $\sigma$ should be set to $0$. However, it is necessary to set $\sigma$ to an appropriate finite value because errors are included in the time-series dataset and the training results of DNN. 
Such $\sigma$ affected by noise cannot be set in advance. 
%また，座標変換に沿った大域的なフラットなランドスケープを探索する必要があるが，%これも困難である．
In addition, the target distributions in this study are assumed to be the global flat minima, because the same $E_{\rm samp}$ surface following the invariant transformation exists. 
Generally, such a target distribution needs an enormous amount of time to sample. 
%そこで，これらを解決可能なサンプリング法として，
%複数の$\sigma$で独立にサンプリングしながら，
%お互いの情報を定期的に交換することで効率的なサンプリングを行う
%レプリカ交換モンテカルロ法を用いた．
Therefore, in this study, we use the replica-exchange Monte Carlo (REMC) method~\cite{hukushima1996exchange} as a sampling method to overcome these problems. 
Such a method enables us to perform efficient sampling by parallel sampling with different noise intensities of $\sigma$ while exchanging noise intensities with each other. 
In the state of a large noise, we can realize global sampling from the abstract distribution
\begin{eqnarray}
\label{samp_func}
{\rm P}'(a_{11},a_{12},a_{21},\cdots,a_{2d\:2d}) = \frac{1}{Z'} \exp\left[-\frac{N}{2\sigma'^2} E_{\rm samp}(a_{11},a_{12},a_{21},\cdots,a_{2d\:2d})\right],
\end{eqnarray}
where $\sigma' > \sigma$. By exchanging this sampling information with the state of a small noise, we can perform efficient sampling from the target distribution $P(a_{11},a_{12},a_{21},\cdots,a_{2d\:2d})$. 
%サンプリング法のパラメータは，
%Aの理論に従い，Bと同様の手順で行った．
The detailed explanation of REMC method and the setting parameters of the method are described in Appendix~\ref{remc_params}, and the target $\sigma$ is determined by analyzing the sampling results as described in Appendix~\ref{appendix_sec3}. 
The procedure of Method 1 is summarized in Algorithm \ref{alg1}.\par
Note that there is no description of how to train a DNN in this study. 
In the training of the deep autoencoder, the number of nodes in the hidden layer is an important hyperparameter. 
On the other hand, since this is a quantity that determines how much the phenomenon is to be reduced, it is considered to be provided by the physicist.

\begin{algorithm}[H]  
\caption{Estimation of the invariant transformation set}         
\label{alg1}                          
\begin{algorithmic}
\item[{\bf Input}:] dataset $D = \left\{\textit{\textbf{q}}_{t_i}^i, \textit{\textbf{p}}_{t_i}^i,\textit{\textbf{q}}_{t_i+\Delta t}^i, \textit{\textbf{p}}_{t_i+\Delta t}^i\right\}_{i=1}^N$ in a given coordinate system.
\item[{\bf Output}:] Invariant transformation set $D_{a} = \{(a_{11},a_{12}\cdots ,a_{1d},a_{21} \cdots ,a_{2d\:2d})_{n_a}\}_{n_{a}=1}^{N_a}$. 
\item[{\bf Step 1}:] Train the deep autoencoder with dataset $D$. 
\item[{\bf Step 2}:] Using the trained deep autoencoder and REMC method, sampling transformation parameters $a_{11},a_{12},a_{21},\cdots,a_{2d\:2d}$ from multiple probability distributions $P'(a_{11},a_{12},a_{21},\cdots,a_{2d\:2d})$ corresponding to different noise intensities $\sigma'$. 
\item[{\bf Step 3}:] Select $\sigma'$ from the distribution structure of the sampling results and output the sampling result of the selected $\sigma'$ state as $D_a$. 
\end{algorithmic}
\end{algorithm}

\subsection{Method 2: Inferring the conservation law from obtained symmetry}
\label{sec_consest}
From the $N_a$ sampling results $D_{a} \coloneqq \{(a_{11},a_{12}\cdots ,a_{1d},a_{21} \cdots ,a_{2d\:2d})_{n_a}\}_{n_{a}=1}^{N_a}$ in  Sec.~\ref{sec_extract}, we propose a method of estimating the infinitesimal transformation, which represents the invariance of the Hamiltonian and the equation of motion. \\

%ここで，無限小変換とサンプリング結果の関係性について述べる．
%ネーターの定理で対象とする連続対称性はリー群をなす．
%The representation of the Lie group is expressed as 
%リー郡の表現は，滑らかなパラメータ集合$\boldsymbol{\theta} = \{\theta_k\}_{k=1}^p$を用いて，
%$
%  A_{ij}(\boldsymbol{\theta}) = a_{ij}(\boldsymbol{\theta})
%$. 
%と表現される．
The set of invariant transformation $M_{\rm invariant}$ is characterized by the $d_{\theta}$-dimensional continuous parameter $\boldsymbol{\theta}$. 
Therefore, $M_{\rm invariant}$ is a $d_{\theta}$-dimensional differential manifold. 
Note that $M_{\rm invariant}$ forms a Lie group as we mentioned in Sec.~\ref{sec_noether}. 
The infinitesimal transformation is estimated as the tangent vector of $M_{\rm invariant}$ at $\boldsymbol{\theta} = \boldsymbol{0}$. 
Using $A(\boldsymbol{\theta})$, we estimate $M_{\rm invariant}$ as
\begin{equation}
    M_{{\rm invariant}} \sim \left. \left\{A(\boldsymbol{\theta}) \left(
\begin{array}{c}
  {\textit{\textbf{q}}}\\  {\textit{\textbf{p}}}
\end{array}
\right)
\right|\:\boldsymbol{\theta} \in \mathbb{R}^{d_{\theta}}\right\}.
\end{equation}
By serializing the transformation matrix $A(\boldsymbol{\theta})$, we define the vector 
\begin{equation}
A'(\boldsymbol{\theta}) = (a'_1(\boldsymbol{\theta}),\cdots,a'_{d'}(\boldsymbol{\theta})) \coloneqq (a_{11}(\boldsymbol{\theta}),\cdots,a_{1d}(\boldsymbol{\theta}),a_{21}(\boldsymbol{\theta}),\cdots,a_{2d}(\boldsymbol{\theta}),\cdots,a_{d1}(\boldsymbol{\theta}),\cdots,a_{2d\:2d}(\boldsymbol{\theta})),
\end{equation}
where $d' = 4d^2$. 
%ここで，この変換行列の要素を並べたベクトル，$A'(\boldsymbol{\theta}) = (a'_1(\boldsymbol{\theta}),\cdots,a'_{d'}(\boldsymbol{\theta})) = (a_{11}(\boldsymbol{\theta}),\cdots,a_{1d}(\boldsymbol{\theta}),a_{21}(\boldsymbol{\theta}),\cdots,a_{2d}(\boldsymbol{\theta}),\cdots,a_{d1}(\boldsymbol{\theta}),\cdots,a_{dd}(\boldsymbol{\theta}))$\\,$d' = d^2$を考える．
%リー群は$A'$の$\boldsymbol{\theta}$を動かすことによって定義される$p$次元可微分多様体と対応する．
%独立な保存則の数と，ベクトルθの次元は一致する（Sec.）．
%今，この多様体の陰関数表現が以下で定義されたとする．
The implicit function representation of the manifold $M_{\rm invariant}$ is defined as 
\begin{eqnarray}
\label{impeqs}
  \begin{cases}
&f_1(a'_1,\cdots,a'_{d'}) = 0\\
&\:\:\:\:\:\:\:\:\:\: \vdots\\
&f_{d'-d_{\theta}}(a'_1,\cdots,a'_{d'}) = 0\\
  \end{cases}.
\end{eqnarray}
%我々が知りたいのは，無限小変換である．
In the representation of the implicit function, the infinitesimal transformation is estimated as the tangent vector of the manifold $M_{{\rm invariant}}$ at the position 
\begin{eqnarray}
e_{\textit{\textbf{I}}} = (\cdots,a_{ij}=0,\cdots,a_{i\:i}=1,\cdots),
\end{eqnarray}
where $i\neq j$ and $e_{\textit{\textbf{I}}}$ is the representation of the identity matrix $\textit{\textbf{I}}$ in the $A'(\boldsymbol{\theta})$ space. 
%この接空間を前節で得られたサンプリング結果から推定する．
We estimate this tangent space ${\rm T}_{\textit{\textbf{I}}} M_{{\rm invariant}} = {\rm T}_{e_{\textit{\textbf{I}}}} M_{{\rm invariant}}$ from the sampling results $D_a$ obtained in Sec.~\ref{sec_extract}.\par
%今，$A'$空間中のある$p$個の変数が構成する部分空間$(b_1,b_2,\cdots,b_p)\subset A'$について，$f_k$のヤコビアン行列
The Jacobian matrix of $f_k$ for parameters of the subset $A'$, $(b_1,b_2,\cdots,b_{d_{\theta}})\subset A'$, is defined as $J_{kl} = \frac{\partial{f_k(a'_1,\cdots,a'_{d'})}}{\partial b_l}$. 
%が$E'$で正則となるとする．
If the Jacobian matrix at $A' = e_{\textit{\textbf{I}}}$ becomes nonsingular, from the implicit function theorem, variables other than $(b_1,b_2,\cdots,b_{d_{\theta}})$, $\{c_{k}\}_{k=1}^{d'-{d_{\theta}}} \coloneqq  A' \setminus \{b_{l}\}_{l=1}^{d_{\theta}}$, can be expressed as 
$c_k = g_{i}(b_1,\cdots,b_{d_{\theta}})$. 
%と表現できる．
%より緩和された見方をすると，$E'$でのヤコビアンが正則な場合，$E'$周辺での多様体の方程式は，以下の$d'-p$個の連立方程式に分解できる．
This implies that, around $e_{\textit{\textbf{I}}}$, the implicit equations in Eq.~\eqref{impeqs} representing the manifold $M_{{\rm invariant}}$ can be decomposed into the following $d'-d_{\theta}$ simultaneous equations:
\begin{eqnarray}
  \begin{cases}
  \label{simul_eq}
&h_{1}(c_1,b_1,\cdots,b_{d_{\theta}})=0\\
&\:\:\:\:\:\:\:\:\:\: \vdots\\
&h_{d'-{d_{\theta}}}(c_{d'-d_{\theta}},b_1,\cdots,b_{d_{\theta}})=0\\
  \end{cases},
\end{eqnarray}
%この方程式を点$E'$の周りで$b_l$について微分すると，
%偏微分を含んだ$d'-p$個の連立方程式
where $b_l$ corresponds to the continuous parameter $\theta_l$ of the continuous transformation $\left[\mathbf{\mathcal{Q}}(\textit{\textbf{q}},\textit{\textbf{p}}, \boldsymbol{\theta}),\mathbf{\mathcal{P}}(\textit{\textbf{q}},\textit{\textbf{p}}, \boldsymbol{\theta})\right]$. 
Differentiating these equations with respect to $b_l$ around a point $e_{\textit{\textbf{I}}}$ yields $d'-d_{\theta}$ simultaneous partial differential equations,
\begin{eqnarray}
\label{diff_eq}
  \begin{cases}
\label{partial_eq}
&\frac{\partial}{\partial b_l}h_{1}(c_1,b_1,\cdots,b_{d_{\theta}})|_{A'=e_{\textit{\textbf{I}}}} = 0\\
&\:\:\:\:\:\:\:\:\:\: \vdots\\
&\frac{\partial}{\partial b_l}h_{d'-d_{\theta}}(c_{d'-d_{\theta}},b_1,\cdots,b_{d_{\theta}})|_{A'=e_{\textit{\textbf{I}}}} = 0\\
  \end{cases}.
\end{eqnarray}
%が得られる．
%この連立微分方程式を解くと，$A'$によって構成さえれる多様体の$E'$周りでの$b_l$方向の接線ベクトル$T_l = (\frac{\partial a_1}{\partial  b_l},\cdots,\frac{\partial  b_l}{\partial  b_l}=1,\cdots,\frac{\partial  b_l'}{\partial  b_l}=0,\cdots,\frac{\partial  a_{d'}}{\partial  b_l})$($l\neq l'$)が得られる．
Solving these simultaneous partial differential equations gives the tangent vector $\left.\frac{A'(b_l)}{\partial b_l}\right|_{A'=e_{\textit{\textbf{I}}}}$ of the manifold around $e_{\textit{\textbf{I}}}$. 
Using the tangent vector as the nonserialized representation $\left.\frac{A(b_l)}{\partial b_l}\right|_{A=\textit{\textbf{I}}}$, we can estimate an infinitesimal transformation as
\begin{equation}
\label{infest}
\left(
\begin{array}{c}
 \delta {\textit{\textbf{q}}}_l\\ \delta {\textit{\textbf{p}}}_l
\end{array}
\right)
 = \varepsilon \left.\frac{A(b_l)}{\partial b_l}\right|_{A=\textit{\textbf{I}}}
 \left(
\begin{array}{c}
  {\textit{\textbf{q}}}\\  {\textit{\textbf{p}}}
\end{array}
\right)
 = \varepsilon \left(
    \begin{array}{ccc}
      \frac{\partial a_{1\:1}}{\partial b_l}|_{A=\textit{\textbf{I}}} & \cdots & \frac{\partial a_{2d\:1}}{\partial b_l}|_{A=\textit{\textbf{I}}}\\
      \vdots & \ddots & \vdots \\
      \frac{\partial a_{1\:2d}}{\partial b_l}|_{A=\textit{\textbf{I}}} & \cdots & \frac{\partial a_{2d\:2d}}{\partial b_l}|_{A=\textit{\textbf{I}}}\\
    \end{array}
  \right) 
 \left(
\begin{array}{c}
  {\textit{\textbf{q}}}\\  {\textit{\textbf{p}}}
\end{array}
\right).
\end{equation} 

\par
%無減少変換はeiでのD_aの接平面と理解される．
%従って，ei近傍のc_kをb_lの一次項で回帰できれば，必要十分な精度の保存則の推定が実現される．
Thus, the invariant transformation is obtained as the tangent vector of the manifold $M_{{\rm invariant}}$ at point $e_{\textit{\textbf{I}}}$. 
Therefore, if $c_k$ can be regressed around $e_{\textit{\textbf{I}}}$ as the first-order polynomial of $\{b_l\}_{l=1}^{d_{\theta}}$, the conservation law can be inferred without approximation. 
%これは，最悪の場合，無限次までの多項式近似が必要となる，ハミルトニアンの推定と比較した際の，保存料推定のアドバンテージである．
Compared with the Hamiltonian estimation and conservation law estimation, this is the advantage of conservation law estimation, because, in general, the Hamiltonian estimation requires infinite-order polynomial approximation. 
%一方で，ノイズのある有限のデータから，ei近傍の接平面を推定することは困難である．これを回避する方法については，考察で議論を行う．
On the other hand, the estimation accuracy of the tangent space ${\rm T}_{e_{\textit{\textbf{I}}}} M_{{\rm invariant}}$ from finite data with noise is often low. 
In this study, we propose a method of estimating the infinitesimal transform with high accuracy by using all sampled transformation data, not only data around $e_{\textit{\textbf{I}}}$. 
Another way to avoid this problem is also discussed in Sec.~\ref{summary}.\par
%そこで本研究では，ei近傍以外の全ての変換のサンプリングデータも使用することで，無限小変換を精度良く推定する方法を提案する．

%今，前節の手法によって$L$個のサンプリング結果$D = \{a_{1}^{'l},a_{2}^{'l},\cdots,a_{d'}^{'l}\}_{l=1}^L$が得られているとする．
The simultaneous equations in Eq.~(\ref{simul_eq}) can be estimated by the following procedure. 
%まず多様体の次元$p$は，多様体の次元推定法などを用いて推定する．
First, the upper limit of the dimension of the manifold $M_{{\rm invariant}}$ is estimated by applying principal component analysis and the ‘‘elbow'' method to $D_a$ as described in~\cite{ulfarsson2008dimension}. 
%あるいは，多様体の次元推定法を用いれば，おおよその多様体の次元を見積もることもできる．これらの手法を用いて，多様体の次元の候補を用意する．
Alternatively, the approximate dimension of $M_{{\rm invariant}}$ can be estimated by using the manifold dimension estimation method such as the method described in~\cite{levina2005maximum}. 
Using such an estimated dimension of $M_{{\rm invariant}}$, we can prepare candidate dimension $d_{\theta}'$. 
%次に，適当な$p' < p_{\rm max}$個の変数セット$(b_1,b_2,\cdots,b_{p'})$を抽出する．
Second, we extract one variable set $(b_1,b_2,\cdots,b_{d_{\theta}'})$. 
%そして，Orthogonal Distance Regressionを用いて$\{c_{k},b_1^l,b_2^l,\cdots,b_{p'}^l\}_{l=1}^{L}$を次のような陰関数
By orthogonal distance regression~\cite{brown1990statistical}, we regress $D_b \equiv \{(c_{k},b_1,b_2,\cdots,b_{d_{\theta}'})_{n_a}\}_{n_a=1}^{N_a}$ to a $d_b$-order implicit polynomial function, 
\begin{eqnarray}
\label{implicit}
\hat{h}_k(c_k, b_1,b_2,\cdots,b_{d_{\theta}'};\beta,\gamma,d_{\theta}') \coloneqq 
\sum_{s_0=0}^{d_b}\sum_{s_1=0}^{d_b}  \cdots \sum_{s_{d_{\theta}'}=0}^{d_b} \gamma_{s_0s_1s_2\cdots s_{d_{\theta}'}} \beta_{s_0s_1s_2\cdots s_{d_{\theta}'}}c_k^{s_0} b_{1}^{s_1}b_{2}^{s_2}\cdots b_{d_{\theta}'}^{s_{d_{\theta}'}}=0, 
\end{eqnarray}
where $\beta$ is the regression coefficient, and $\gamma$ is a binary vector indicating whether the basis is selected. 
%で回帰する．
%回帰モデル$lm$はBICによるモデル選択によって決定する．
%これによって連立方程式Eq.\eqref{simul_eq})が得られる．
The indicator vector $\gamma$ and the dimension of the manifold $d_{\theta}'$ are determined by a model selection method, such as the Bayesian information criterion (BIC)~\cite{schwarz1978estimating}. To select the model, it is necessary to estimate the likelihood. The method of estimating the likelihood is described in Appendix~\ref{est_likeli}. 
If $d_{\theta}\leq 2$, $d_{\theta}'$ can be determined by visualization. 
%ここで，サンプリングデータ全体を回帰することから，ei近傍とは異なり，指数の上限Sは十分に大きくとる必要があることに注意されたい．
Note that, unlike the estimation of the tangent space ${\rm T}_{e_{\textit{\textbf{I}}}} M_{{\rm invariant}}$, the upper limit $d_b$ of the order of polynomial function must be sufficiently large because the entire sampling data is regressed. 
%この回帰とモデル選択を，全てのC_kについてそれぞれ行えば，N_invariantの陰関数表示が得られる．
This regression and model selection is performed for all $c_k$; then, an implicit function representation of $M_{\rm invariant}$ can be obtained. 
\par
%統計的なモデル選択のために，次のような尤度関数を設定する．\par
%これによって連立微分方程式Eq.\eqref{simul_eq})と連立微分方程式Eq.\eqref{diff_simul_eq})が得られる．
From the obtained simultaneous equations, we obtain the simultaneous differential equations. 
%ヤコビアン行列$J_{kl}$が正則でない場合，この方程式の解が発散したり，不定となる．その場合は，変数セット$\{b_1^l,b_2^l,\cdots,b_{p'}^l\}$を再度抽出し直して同様のことを繰り返す．
If the Jacobian matrix $J_{kl}$ is singular, the solution of the simultaneous equations diverges or becomes indefinite. 
In that case, the variable set $\{(b_1, \cdots, b_{d_{\theta}'}) \} $ is extracted again and the same procedure is repeated. 
If the Jacobian matrix $J_{kl}$ is nonsingular, we can obtain the infinitesimal transformation according to Eq.~\eqref{infest}. 
%この手法は，使用するサンプリングデータをei近傍に絞っていくことで，より低い次数での多項式回帰で，無減少変換の推定が可能となる．
In this method, by narrowing down the regressing area of $D_a$ to the neighborhood of $e_{\textit{\textbf{I}}}$, we obtain a higher accurate estimation of infinitesimal transformation with a lower-order polynomial function in Eq.~\eqref{implicit}.

\begin{algorithm}[H]                  
\caption{Estimation of infinitesimal transformation}         
\label{alg2}                          
\begin{algorithmic}
\item[{\bf Input}:] Sampling results of Method 1, $D_{a} = \{(a_{11},a_{12}\cdots ,a_{1d},a_{21} \cdots ,a_{2d\:2d})_{n_a}\}_{n_{a}=1}^{N_a}$, and $d_{\theta}'$.
\item[{\bf Output}:] Infinitesimal transformation, $\delta {\textit{\textbf{q}}}_l$, $\delta {\textit{\textbf{p}}}_l$. 
\item[{\bf Step 1}:] Extract $D_b = \{(c_{k},b_1,b_2,\cdots,b_{d_{\theta}'})_{n_a}\}_{n_a=1}^{N_a}$ from $D_a$. 
\item[{\bf Step 2}:] Fit $D_b$ with the implicit polynomial function $\hat{h}_k(c_k, b_1^l,b_2^l,\cdots,b_{d_{\theta}'}^l;\beta,\gamma,d_{\theta}')$ [Eq.~\eqref{implicit}] for each $c_k$. 
\item[{\bf Step 3}:] Estimate the likelihood [Eq.~\eqref{likelihood}] by numerical integration of $Z$ [Eq.~\eqref{normalizeconstant}]. 
\item[{\bf Step 4}:] Select the indicator vector $\gamma$ and the dimension $d_{\theta}'$ of $M_{\rm invariant}$ in Eq.~\eqref{implicit} for each $c_k$ using the BIC, . 
\item[{\bf Step 5}:] Determine whether the Jacobi matrix $J_{kl} = \frac{\partial h_k(c_{k},b_1,\cdots,b_{d_{\theta}})}{\partial b_l}$ is nonsingular. If $J_{kl}$ is singular, return to Step 1 and re-extract $D'_b$. 
\item[{\bf Step 6}:] Differentiate the obtained simultaneous equations with respect to $b_l$ around a point $e_{\textit{\textbf{I}}}$ to obtain Eq.~\eqref{partial_eq}. 
\item[{\bf Step 7}:] Solve the simultaneous equations in Eq.~\eqref{partial_eq} and obtain the infinitesimal transformation, $\delta {\textit{\textbf{q}}}_l$, $\delta {\textit{\textbf{p}}}_l$.
\end{algorithmic}
\end{algorithm}

\section{Results}
\label{results}
We evaluate the proposed method using one geometrical structure and three physical systems: 
(\rnum{1}) a half sphere, (\rnum{2}) constant-velocity linear motion, (\rnum{3}) a two-dimensional central force system, and (\rnum{4}) a collective motion system. 
%ケースa)は，回転対称性のあるケースで，手法1によってその対称性に対応した変換の集合が獲得できるかを確認する．
%ケースb)及びc)はそれぞれ，運動量保存則，角運動量保存則が成り立つことがわかっている系であり，これによって手法2の検証を行う．
%その上で，複雑なcollective motion系であるd)に手法を適用し，複雑な大自由度系の保存則推定を試みる．
Case~(\rnum{1}) has a rotational symmetry. In case~(\rnum{1}), we confirm that Method 1 can obtain a set of transformations corresponding to the symmetry.
Cases~(\rnum{2}) and (\rnum{3}) are systems that conserve the momentum and angular momentum, respectively. Using these cases, we verified Method 2. 
Finally, we apply both proposed methods to (\rnum{4}), which is a complicated collective motion system, and attempted to infer the collective coordinate and conservation law. In each case, the parameters of DNN are set as described in Appendix~\ref{dnn_params}, and REMC are set as described in Appendix~\ref{remc_params}.  
\subsection*{(\rnum{1}) Half sphere}
\begin{figure}[htbp]
  \begin{center}
   \includegraphics[width=150mm]{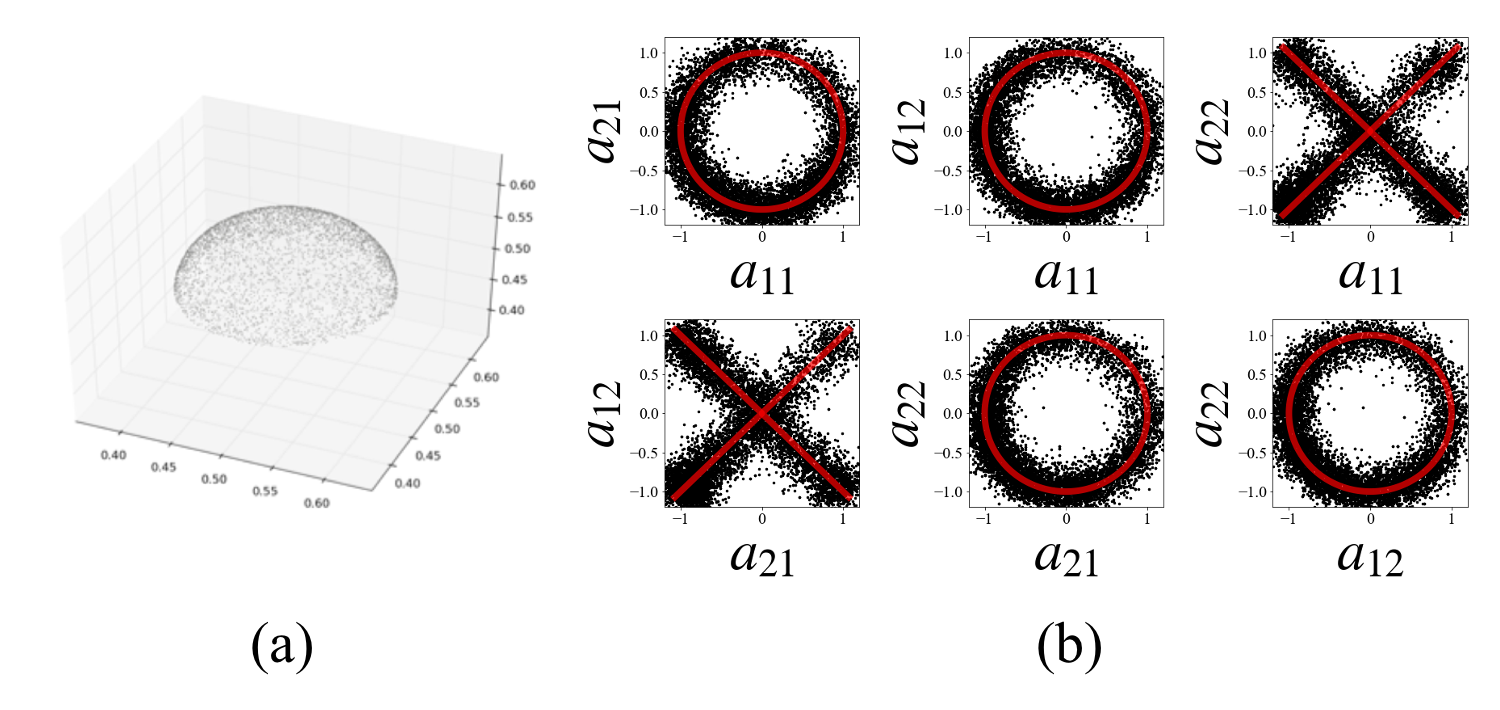}
  \caption{Results of case~(\rnum{1}): half sphere. (a) Dataset using the evaluation. There are 1,671 samples. 
(b) Black dots represent sampling distributions obtained by Method 1 and red curves represent fitting curves estimated by Method 2. Each graph shows six combinations of four transformation variables $a_{ij}$.}
  \label{fig_result1}
  \end{center}
\end{figure}

The dataset of case~(\rnum{1}) was generated by the function
\begin{eqnarray}
\label{eqa}
x_1^2+x_2^2+x_3^2 = r,\:\: (x_3 > 0),
\end{eqnarray}
where $r$ was set to be $0.25$. 
We generated 1,671 samples according to Eq.~\eqref{eqa}. 
The dataset of case~(\rnum{1}) [shown in Fig.~\ref{fig_result1}(a)] was used to verify the ability of Method 1 described in Sec.~\ref{sec_extract}, which extracts the symmetry. 
We set the coordinate system as $(x_1, x_2, x_3)$ and limit the transformation on the $x_1$-$x_2$ plane. 
In such a case, the transformation matrix $A$ is defined as
\begin{equation}
\label{r1a}
A = 
\left(
    \begin{array}{ccc}
      a_{11} & a_{21} & 0\\
      a_{12} & a_{22} & 0\\
      0 & 0 & 0\\
    \end{array}
  \right). 
\end{equation} 
In this coordinate system, the half sphere has a rotation symmetry and a mirror symmetry. The rotation symmetry transformation is represented as
\begin{equation}
A_{\rm rot}(\theta_{\rm rot}) = 
\left(
    \begin{array}{cc}
      \cos(\theta_{\rm rot}) & \sin(\theta_{\rm rot})\\
      -\sin(\theta_{\rm rot}) & \cos(\theta_{\rm rot})\\
    \end{array}
  \right), 
\end{equation} 
where $\theta_{\rm rot}$ is a rotation angle, and the mirror symmetry transformation is represented as 
\begin{equation}
A_{\rm mirror}(\theta_{\rm mirror}) = 
\left(
    \begin{array}{cc}
      \cos(2\theta_{\rm mirror}) & \sin(2\theta_{\rm mirror})\\
      \sin(2\theta_{\rm mirror}) & -\cos(2\theta_{\rm mirror})\\
    \end{array}
  \right), 
\end{equation}
where $\theta_{\rm mirror}$ is an angle of the mirror plane with the $x_1$ axis. 
The mirror symmetry is a discrete symmetry; therefore, the invariant transformation of the half sphere is represented as
$A_{\rm tot}(\theta_{\rm rot}, \theta_{\rm mirror}) \coloneqq A_{\rm rot}(\theta_{\rm rot}) [A_{\rm mirror}(\theta_{\rm mirror})]^m$, where $m \coloneqq \{0,1\}$ and 
\begin{eqnarray}
\label{r1r}
A_{\rm rot}(\theta_{\rm rot}) [A_{\rm mirror}(\theta_{\rm mirror})]^0 &\coloneqq& A_{\rm rot}(\theta_{\rm rot}),\\
\label{r1m}
A_{\rm rot}(\theta_{\rm rot}) [A_{\rm mirror}(\theta_{\rm mirror})]^1 &\coloneqq& 
\left(
    \begin{array}{cc}
      \cos(2\theta_{\rm mirror} - \theta_{\rm rot}) & \sin(2\theta_{\rm mirror} - \theta_{\rm rot})\\
      \sin(2\theta_{\rm mirror} - \theta_{\rm rot}) & -\cos(2\theta_{\rm mirror} - \theta_{\rm rot})\\
    \end{array}
  \right) = A_{\rm mirror}\left(\theta'\right),\\
  \theta' &\coloneqq& \theta_{\rm mirror} - \frac{\theta_{\rm rot}}{2}.
\end{eqnarray}
By comparing Eq.~\eqref{r1a} with Eqs.~\eqref{r1r} and \eqref{r1m}, we obtain the implicit function representation of the invariant transformation $A_{\rm tot}(\theta_{\rm rot}, \theta_{\rm mirror})$ as
\begin{eqnarray}
  \begin{cases}
a_{11}^2 + a_{21}^2 = 1\\
a_{11}^2 + a_{12}^2 = 1\\
(a_{11} + a_{22})(a_{11} - a_{22}) = a_{11}^2 - a_{22}^2 = 0\\
(a_{21} - a_{12})(a_{21} + a_{12}) = a_{21}^2 - a_{12}^2 = 0\\
a_{21}^2 + a_{22}^2 = 1\\
a_{12}^2 + a_{22}^2 = 1 
\end{cases}.
\end{eqnarray}
Method 1 was applied to such a system. \par
The sampling results of $a_{ij}$ are shown in Fig.~\ref{fig_result1}(b) as black dots. 
In the figures, the red curves were fitted by the selected implicit polynomial functions using the BIC. 
%In all three cases, the dimensions of the manifold representing each Lie group was estimated to be 1 using PCA and the ‘‘elbow'' method. 
The fitting results are 
\begin{eqnarray}
  \begin{cases}
a_{11}^2 + 0.99a_{21}^2 = 1\\
a_{11}^2 + a_{12}^2 = 1\\
a_{11}^2 - a_{22}^2 = 0\\
a_{21}^2 - a_{12}^2 = 0\\
a_{21}^2 + a_{22}^2 = 1\\
a_{12}^2 + a_{22}^2 = 1 
\end{cases},
\end{eqnarray}
where we determine $d_{\theta}'$ to be 1 by visualizing the distribution of $D_a$. 

\subsection*{(\rnum{2}) Constant-velocity linear motion}
\begin{figure}[t]
  \begin{center}
   \includegraphics[width=160mm]{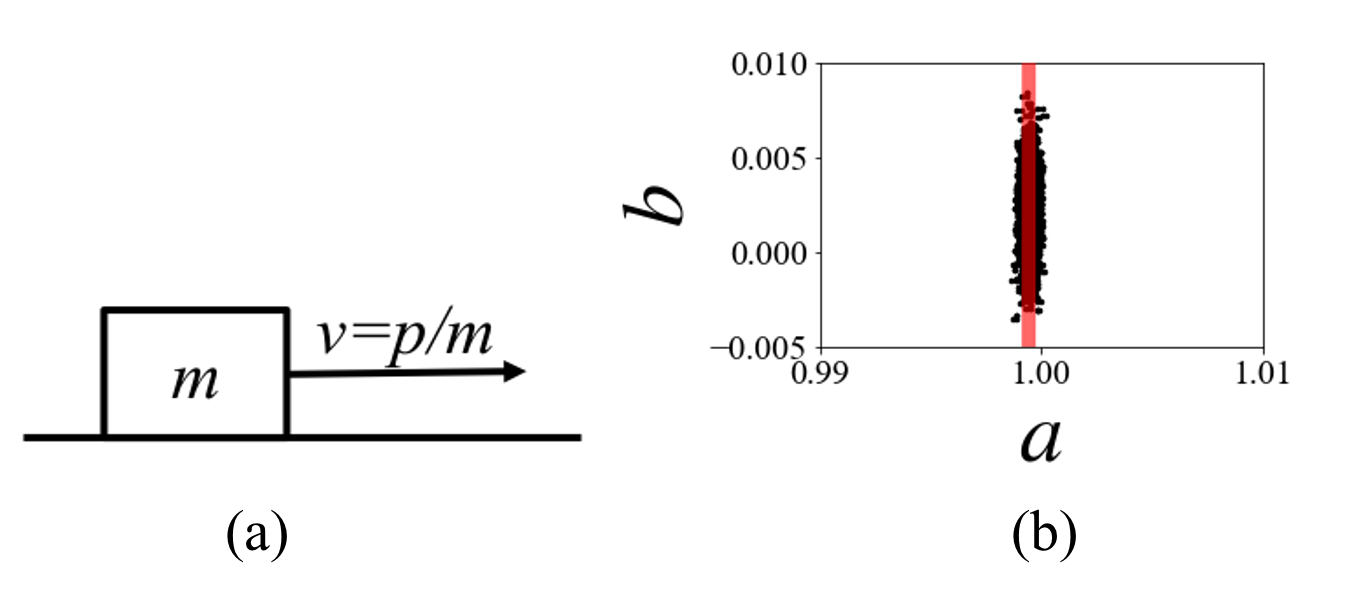}
  \caption{Results of case~(\rnum{2}): constant-velocity linear motion. (a) Conceptual diagram of constant-velocity linear motion. 
(b) Black dots represent sampling distribution obtained by Method 1 and the red line represents the fitting curve estimated by Method 2.}
  \label{fig_result2}
  \end{center}
\end{figure}

The dataset of case~(\rnum{2}) was generated using the one-dimensional Hamiltonian system 
\begin{eqnarray}
\label{eqb}
H_2 = \frac{p^2}{2m},
\end{eqnarray}
where $m$ was set to be $1$. We generated 1,000 samples by solving Eq.~\eqref{eqb}. 
In this case, we show that the proposed method can infer the momentum conservation law. 
We set the coordinate system as $(q,1,p,1)$. 
In such a coordinate system, $q$ and $p$ are related as $p = m\dot{q}$. This means that $q$ and $p$ have the same coordinate transformation. Thus, the transformation matrix $A$ is defined as
\begin{equation}
\label{sampab}
A =\left(
    \begin{array}{cccc}
      a & b & 0 & 0\\
      0 & 1 & 0 & 0\\
      0 & 0 & a & 0\\
      0 & 0 & 0 & 1\\
    \end{array}
  \right).
\end{equation}
As a result, two parameters $a$ and $b$ must be sampled. 
The sampling results of $a_{ij}$ are shown in Fig.~\ref{fig_result2} as black dots and the red curves were fitted by the selected implicit polynomial function using the BIC. 
%In all three cases, the dimensions of the manifold representing each Lie group was estimated to be 1 using PCA and the ‘‘elbow'' method. 
The fitting result is  
\begin{eqnarray}
a=1.0.
\end{eqnarray}
The simultaneous partial differential equations in Eq.~\eqref{diff_eq}, where $b_l = b$, were obtained from the fitting results. From the solution of the simultaneous partial differential equations, we obtained the infinitesimal transformation
\begin{eqnarray}
  \begin{cases}
  \label{result_linear}
\delta q = \epsilon \frac{\partial a}{\partial b} q + \epsilon \frac{\partial b}{\partial b} = \epsilon\\ 
\delta p = \epsilon \frac{\partial a}{\partial b} p = 0 
  \end{cases},
\end{eqnarray}
where we determined $d_{\theta}'$ to be $1$ by visualizing the distribution of $D_a$. 
By substituting this Eq.~\eqref{result_linear} into Eq.~\eqref{noether} and solving the equations, we estimated the conserved value $G_{\delta}$ as $G_{\delta} = 1.0\varepsilon p$. 
This result shows that the momentum $p$ was conserved. 

\subsection*{(\rnum{3}) Two-dimensional central force system}
The dataset of case~(\rnum{3}) was generated using the Hamiltonian system
\begin{eqnarray}
\label{eqc}
H_3 = \frac{1}{2m} {\textit{\textbf{p}}}^2 + G\frac{mM}{\left|{\textit{\textbf{q}}}\right|},
\end{eqnarray}
where ${\textit{\textbf{q}}}\coloneqq(q_1,q_2)$, ${\textit{\textbf{p}}}\coloneqq(p_1,p_2)$, and $m$, $M$, and $G$ were set to be 1. 
We generated 1,000 samples by solving Eq.~\eqref{eqc}. 
We set the coordinate system as $(q_1,q_2,p_1,p_2)$. 
In such a coordinate system, $q_j$ and $p_j$ are related as $p_j = m\dot{q_j}$. Thus, $q_j$ and $p_j$ have the same coordinate transformation, and the transformation matrix $A$ is defined as
\begin{equation}
\label{transc}
A  =
  \left(
    \begin{array}{cccc}
      a_{11} & a_{21} & 0 & 0\\
      a_{12} & a_{22} & 0 & 0 \\
      0 & 0 & a_{11} & a_{21}\\
      0 & 0 & a_{12} & a_{22}\\
    \end{array}
  \right). 
\end{equation}
As a result, only four parameters $a_{ij}$ must be sampled.\par
In the canonical dynamics of $H_3$, it is impossible to transform one orbit to another with the same energy but different long-axis radii using the linear transformation $A$ in Eq.~\eqref{transc}. 
Therefore, the invariant transformation for $S_i$ can be represented as a product of invariant transformation for subspace $S_i^{{\rm radii}}$ for specific energy and long-axis radii. 
This implies that the invariant transformation set for certain energy $E_i$ and certain long-axis radii must include some invariant transformations for $S_i$. 
On the basis of this property of the Hamilton system of $H_3$, we apply the proposed method only to the time-series dataset of a circular orbit with radius 1.\par
\begin{figure}[tp]
  \begin{center}
   \includegraphics[width=160mm]{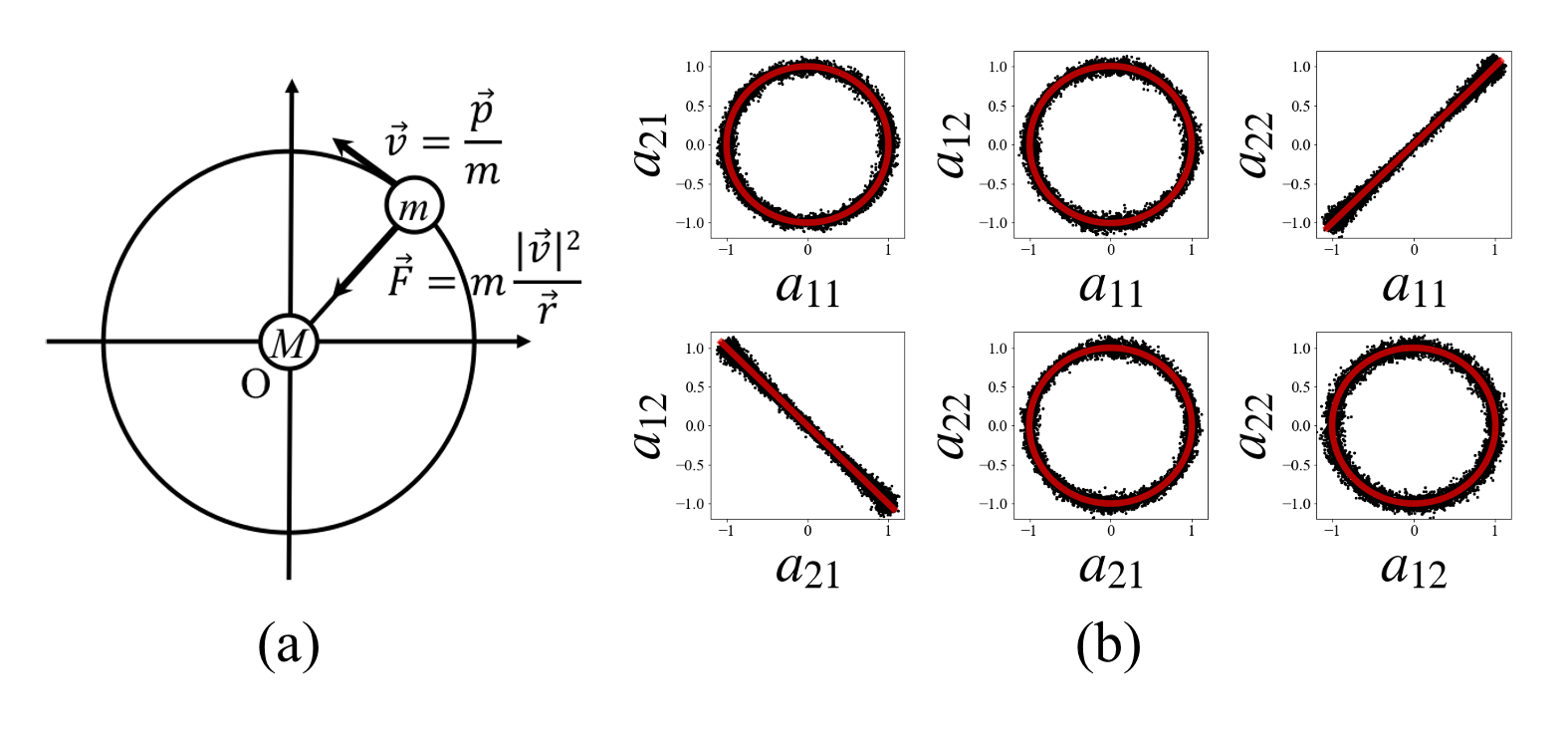}
  \caption{Results of case~(\rnum{3}): two-dimensional central force system. (a) Conceptual diagram of two-dimensional central force system. 
(b) Black dots represent sampling distributions obtained by Method 1 and red curves represent fitting curves estimated by Method 2. Each graph shows six combinations of four transformation variables $a_{ij}$.}
  \label{fig_result3}
  \end{center}
\end{figure}

The sampling results of $a_{ij}$ are shown in Fig.~\ref{fig_result3} as black dots and the red curve in each figure was fitted by the selected polynomial function using the BIC. 
%In all three cases, the dimensions of the manifold representing each Lie group was estimated to be 1 using PCA and the ‘‘elbow'' method. 
The fitting results are 
\begin{eqnarray}
  \begin{cases}
a_{11}^2 + 0.99a_{21}^2 = 1\\
a_{11}^2 + 0.98a_{12}^2 = 1\\
a_{11} - a_{22} = 0\\
a_{21} + 0.99a_{12} = 0\\
a_{21}^2 + 1.01a_{22}^2 = 1.01\\
a_{12}^2 + 1.02a_{22}^2 = 1.02\\
\end{cases},
\end{eqnarray}
where we determine $d_{\theta}'$ to be $1$ by visualizing the distribution of $D_a$. 
The simultaneous partial differential equations in Eq.~\eqref{diff_eq}, where $b_l = a_{21}$, were obtained from the fitting results. By solving the simultaneous partial differential equations, we obtained the infinitesimal transformation
\begin{eqnarray}
\delta \textit{\textbf{q}} &=&
\varepsilon 
\left(
    \begin{array}{cc}
      \frac{\partial a_{11}}{\partial a_{21}} & \frac{\partial a_{21}}{\partial a_{21}}\\
      \frac{\partial a_{12}}{\partial a_{21}} & \frac{\partial a_{21}}{\partial a_{21}}\\
    \end{array}
  \right)\textit{\textbf{q}} 
  =
\varepsilon 
\left(
    \begin{array}{cc}
\left.\frac{-2\times 0.99a_{21}}{2a_{11}}\right|_{A=\textit{\textbf{I}}} & 1\\
-1/0.99 & \left.\frac{-2a_{21}}{1.01 \times 2a_{22}}\right|_{A=\textit{\textbf{I}}}
    \end{array}
  \right)\textit{\textbf{q}} \\
 &=& 
\left(
    \begin{array}{cc}
0 & \varepsilon \\
-1.01\varepsilon & 0
    \end{array}
  \right)\textit{\textbf{q}} \approx
  \left(
    \begin{array}{cc}
0 & \varepsilon\\
-\varepsilon & 0
    \end{array}
  \right)\textit{\textbf{q}}, \label{result_orbitq}\\ 
\delta \textit{\textbf{p}} &=&
\varepsilon     \left(\begin{array}{cc}
      \frac{\partial a_{11}}{\partial a_{21}} & \frac{\partial a_{21}}{\partial a_{21}}\\
      \frac{\partial a_{12}}{\partial a_{21}} & \frac{\partial a_{21}}{\partial a_{21}}\\
    \end{array}\right)\textit{\textbf{p}}
\approx \left(
    \begin{array}{cc}
0 & \varepsilon\\
-\varepsilon & 0
    \end{array}
  \right)\textit{\textbf{p}}\label{result_orbitp},
\end{eqnarray}
where the values in the final formula are to one decimal place.
By substituting Eqs.~\eqref{result_orbitq} and \eqref{result_orbitp} into Eq.~\eqref{noether} and solving the equation, we estimated the conserved value $G_{\delta}$ as $G_{\delta} = \varepsilon (x_1p_2 - x_2p_1)$. 
This result shows that the angular momentum was conserved. 

\subsection*{(\rnum{4}) Collective motion system}
In this case, we apply our framework to an $N_{\rm R}$-body collective motion system called the Reynolds boid model~\cite{reynolds1987flocks}. In this model, each individual moves following three forces, which are the force attracting each other, separating each other, and aligning the orientation of each other:
\begin{eqnarray}
\label{eqd}
\frac{d {\textit{\textbf{p}}}_j}{d t} &=&  -W_{\rm att}\left({\textit{\textbf{p}}}_j  - \frac{\sum_{k\in K_{\rm att}}{\textit{\textbf{p}}}_k}{n_{\rm att}}\right) + W_{\rm sep}\left(\sum_{k\in K_{\rm sep}}\frac{({\textit{\textbf{q}}}_j - {\textit{\textbf{q}}}_k)}{|{\textit{\textbf{q}}}_j - {\textit{\textbf{q}}}_k|}\right) + W_{\rm ali}\left({\textit{\textbf{q}}}_j - \frac{\sum_{k\in K_{\rm ali}}{\textit{\textbf{q}}}_k}{n_{\rm ali}}\right),\\
\frac{d {\textit{\textbf{q}}}_j}{d t} &=& {\textit{\textbf{p}}}_j,
\end{eqnarray}

\begin{eqnarray*}
K_{\rm att} &=& \left\{k \:\left|\: |{\textit{\textbf{q}}}_k - {\textit{\textbf{q}}}_j| < r_{\rm att},\: \arccos\left(\frac{{\textit{\textbf{p}}}_k \cdot {\textit{\textbf{p}}}_j}{|{\textit{\textbf{p}}}_k| |{\textit{\textbf{p}}}_j|}\right)<\theta_{\rm att},\:k \neq j\right.\right\},\\
K_{\rm sep} &=& \left\{k \:\left|\: |{\textit{\textbf{q}}}_k - {\textit{\textbf{q}}}_j| < r_{\rm sep},\: \arccos\left(\frac{{\textit{\textbf{p}}}_k \cdot {\textit{\textbf{p}}}_j}{|{\textit{\textbf{p}}}_k| |{\textit{\textbf{p}}}_j|}\right)<\theta_{\rm sep},\:k \neq j\right.\right\},\\
K_{\rm ali} &=& \left\{k \:\left|\: |{\textit{\textbf{q}}}_k - {\textit{\textbf{q}}}_j| < r_{\rm ali},\: \arccos\left(\frac{{\textit{\textbf{p}}}_k \cdot {\textit{\textbf{p}}}_j}{|{\textit{\textbf{p}}}_k| |{\textit{\textbf{p}}}_j|}\right)<\theta_{\rm ali},\:k \neq j\right.\right\},\\
n_{\rm att} &=& \sum_{k\in K_{\rm att}} 1,\:\:n_{\rm ali} = \sum_{k\in K_{\rm ali}} 1,
\end{eqnarray*}
where ${\textit{\textbf{q}}}\coloneqq(q_1,q_2,q_3)$, ${\textit{\textbf{p}}}\coloneqq(p_1,p_2,p_3)$, and $j$, $k$ represent the index of an individual. 
The attraction, separation, and alignment terms are represented by the first, second, and third terms in Eq.~\eqref{eqd}, and each force has the interaction range, $r_{\rm att}$, $r_{\rm sep}$, $r_{\rm ali}$, and angle of view, $\theta_{\rm att}$, $\theta_{\rm sep}$, $\theta_{\rm ali}$, respectively. 
The parameters $W_{\rm att}$, $W_{\rm sep}$, $W_{\rm ali}$, $r_{\rm att}$, $r_{\rm sep}$, $r_{\rm ali}$, $\theta_{\rm att}$, $\theta_{\rm sep}$, and $\theta_{\rm ali}$ of the Reynolds boid model can be tuned to simulate the collective motion of living things such as birds or fish~\cite{reynolds1987flocks, couzin2002collective}. 
In this study, we focused on a parameter set that simulates the torus-type behavior of a school of fish in the sea. Such a torus-type collective motion can be realized in a two-dimensional space. 
Therefore, we set the dimension to two in this study. 
By solving Eq.~\eqref{eqd}, we generated 2,000 steps of time-series data of the torus-type collective motion by 200 individuals.\par
\begin{figure}[htbp]
  \begin{center}
   \includegraphics[width=160mm]{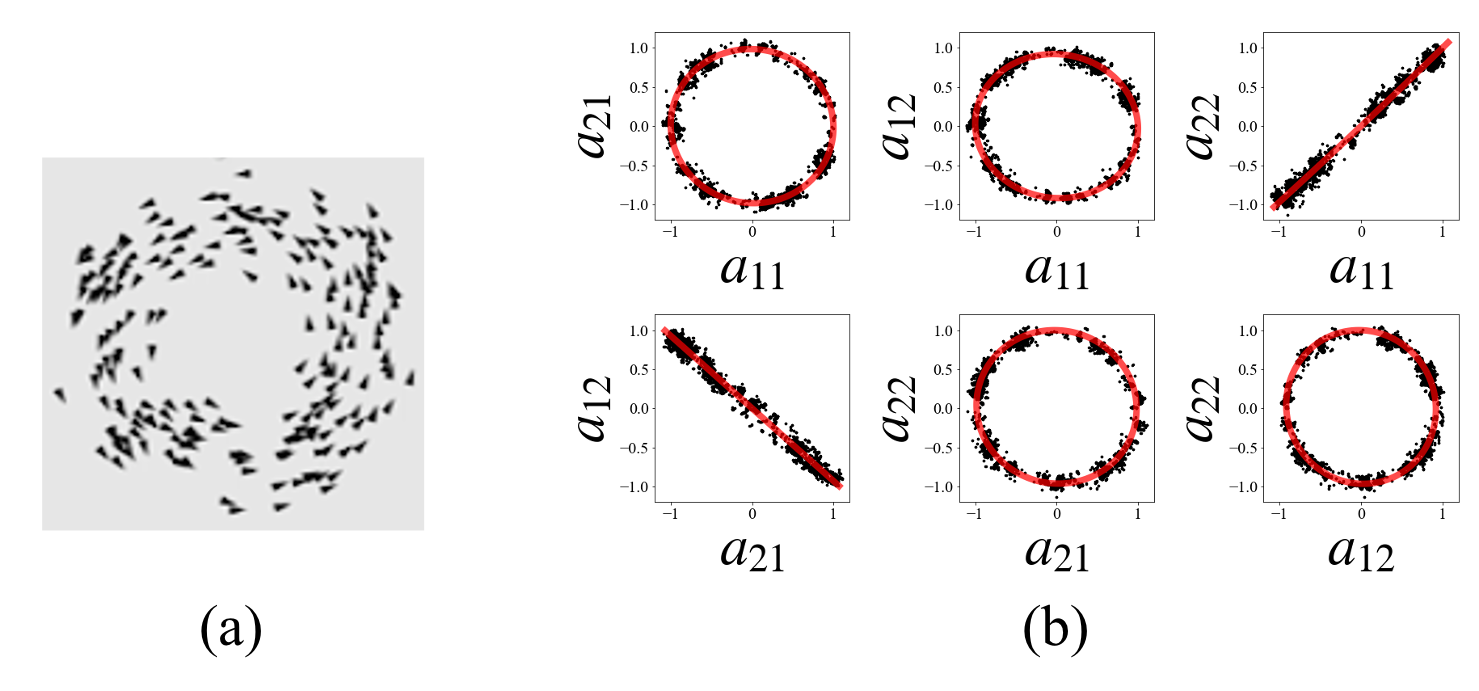}
  \caption{Results of case~(\rnum{4}): collective motion system. (a) Simulation snapshot of torus-type collective motion. The simulation data were applied to the proposed method. 
(b) Black dots represent sampling distributions obtained by Method 1 and red curves represent fitting curves estimated by Method 2. Each graph shows six combinations of four transformation variables $a_{ij}$.}
  \label{fig_result4}
  \end{center}
\end{figure}

To infer the conservation law of collective motion, we need to set a candidate collective coordinate. In this study, we set the collective coordinate on the basis of the following considerations. 
First, from the visual symmetry of the motion, the average position $(\bar{q}_1,\bar{q}_2)$ and average momentum $(\bar{p}_1,\bar{p}_2)$ of all particles over time are set as the origin of the coordinate system. 
Second, since the same behavior is observed regardless of the individual, each individual is considered to have no degree of freedom. 
From these considerations, we set the coordinate system as $(\tilde{\textit{\textbf{q}}},\tilde{\textit{\textbf{p}}}) = (q_1 - \bar{q}_1, q_2 - \bar{q}_2, p_1 - \bar{p}_1, p_2 - \bar{p}_2)$, and prepared the dataset as 
\begin{eqnarray}
D &=& \{\tilde{\textit{\textbf{q}}}(t_i)_i,\tilde{\textit{\textbf{p}}}(t_i)_i,\tilde{\textit{\textbf{q}}}(t_i+\Delta t)_i,\tilde{\textit{\textbf{p}}}(t_i+\Delta t)_i\}_{i = 1}^{N_{\rm R}T}\\
&\coloneqq& \{\tilde{\textit{\textbf{q}}}(t_{jk})_{jk},\tilde{\textit{\textbf{p}}}(t_{jk})_{jk},\tilde{\textit{\textbf{q}}}(t_{jk}+\Delta t)_{jk},\tilde{\textit{\textbf{p}}}(t_{jk}+\Delta t)_{jk}\}_{\langle{}j,k\rangle{}},
\end{eqnarray}
where $N_{\rm R} = 200$, $T = 2,000$, and $\langle{}j,k\rangle{}$ represents all combinations of individuals $j$ and time steps $k$. 
We randomly selected 5,000 samples from this dataset for the training of DNN. 
Then, the transformation matrix $A$ is defined as 
\begin{equation}
A = 
  \left(
    \begin{array}{cccc}
      a_{11} & a_{21} & 0 & 0\\
      a_{12} & a_{22} & 0 & 0\\
      0 & 0 & a_{11} & a_{21}\\
      0 & 0 & a_{12} & a_{22}\\
    \end{array}
  \right). 
\end{equation}
The sampling results of $a_{ij}$ are shown in Fig.~\ref{fig_result4}(b) as black dots and the red curves fitted by the selected implicit polynomial functions using the BIC. 
%In all three cases, the dimensions of the manifold representing each Lie group was estimated to be 1 using PCA and the ‘‘elbow'' method. 
The fitting results of the selected implicit polynomial functions are 
\begin{eqnarray}
\begin{cases}
a_{11}^2 + 1.03a_{21}^2 + 0.039a_{11}a_{21} = 1\\
a_{11}^2 + 1.18a_{12}^2 + 0.077 a_{11}a_{12} = 1\\
a_{11} - 1.016a_{22} + 0.016a_{11}^2 = 0\\
a_{21} + 1.077a_{12} = 0\\
-0.038a_{22} + a_{21}^2 + 1.005a_{22}^2 + 0.051a_{21}a_{22} = 0.967\\
-0.031a_{22} + a_{12}^2 + 0.877a_{22}^2 + 0.056a_{12}a_{22} = 0.845
\end{cases},
\end{eqnarray}
where we determine $d_{\theta}'=1$ by visualizing the distribution of $D_a$. 
The simultaneous partial differential equations in Eq.~\eqref{fig_result4}, where $b_l = a_{12}$, were obtained from the fitting results. By solving the simultaneous equations, we obtained the infinitesimal transformation
\begin{eqnarray}
\delta \textit{\textbf{q}} &=&
\varepsilon \left(
    \begin{array}{cc}
      \frac{\partial a_{11}}{\partial a_{21}} & \frac{\partial a_{21}}{\partial a_{21}}\\
      \frac{\partial a_{12}}{\partial a_{21}} & \frac{\partial a_{21}}{\partial a_{21}}\\
    \end{array}
  \right)\textit{\textbf{q}} = \varepsilon
  \left(
    \begin{array}{cc}
\left.\frac{-1.03\times 2a_{21}-0.039a_{11}}{2a_{11}+0.039}\right|_{A=\textit{\textbf{I}}}&1\\
-1/1.077 & \left.\frac{-2a_{21} - 0.051a_{22}}{1.005 \times 2a_{22} - 0.038 + 0.051a_{12}}\right|_{A=\textit{\textbf{I}}}
    \end{array}
  \right)\textit{\textbf{q}}\\
&=& \left(
    \begin{array}{cc}0.019\varepsilon & \varepsilon\\
    -0.928\varepsilon&0.026\varepsilon\end{array}
  \right)\textit{\textbf{q}} \approx \left(
    \begin{array}{cc}0&\varepsilon\\-\varepsilon&0\end{array}
  \right)\textit{\textbf{q}},\label{result_dq}\\
\delta \textit{\textbf{p}} &=&
\varepsilon \left(
    \begin{array}{cc}
      \frac{\partial a_{11}}{\partial a_{21}} & \frac{\partial a_{21}}{\partial a_{21}}\\
      \frac{\partial a_{12}}{\partial a_{21}} & \frac{\partial a_{21}}{\partial a_{21}}\\
    \end{array}
  \right)\textit{\textbf{p}} \approx \left(
    \begin{array}{cc}0&\varepsilon\\-\varepsilon&0\end{array}
  \right)\textit{\textbf{p}},\label{result_dp}
\end{eqnarray}
where the values in the final formula are to one decimal place. 
By substituting Eqs.~\eqref{result_dq} and \eqref{result_dp} into Eq.~(\ref{noether}) and solving the equation, the conserved value $G_{\delta}$ was estimated as $G_{\delta} = \varepsilon (x_1p_2 - x_2p_1)$. 
This result shows that the angular momentum was conserved. 

\section{Summary and Discussion}
\label{summary}
%a)で手法1の有効性が確認された
%ケースA)の結果から，手法1によって対称性が抽出可能であることが確認された．
From the results of case~(\rnum{1}), we confirm that Method 1 could be used to extract the symmetry. 
%さらに，ケースb),c)において，それぞれの系で期待された運動量保存則と角運動量保存則が保存されていることが推定されたことは，手法2が有効であることを示す．
The results of cases~(\rnum{2}) and (\rnum{3}), wherein the expected conservation laws were inferred, show that Method 2 is effective. 
By comparing cases~(\rnum{1}) and (\rnum{3}), we observe differences in the selected implicit polynomial functions in the $a_{11}$-$a_{22}$ and $a_{21}$-$a_{12}$ spaces. 
These differences emerged from the mirror symmetry in case~(\rnum{1}). 
This finding supports the assertion that the method works well in extracting the symmetry of a system. 
%The proposed method may have the potential to estimate the conservation laws of physical systems for which it has been difficult to obtain the conservation law analytically, such as a Hamiltonian with unknown symmetry and the symmetry of the effective Hamiltonian in a many-body system. 
%d)で，collective coordinateと保存則推定が可能なことが示された
%d)の角運動量保存則は，先行研究でその保存則が示唆されていたので，これと整合性をもつ．
For a more practical collective motion system [i.e., case~(\rnum{4})], we inferred the angular momentum conservation law; the results thereof are consistent with a previous study~\cite{couzin2002collective}. In the previous study, it is suggested that angular momentum is conserved in torus-type swarming patterns. 
Additionally, the finding of a conservation law in the collective coordinates, where the degree of freedom of an individual was degenerated and the origin of the coordinates is the average position and momentum of the swarm, suggests that a dynamical system with a large degree of freedom can be reduced to a central force dynamical system.\par
%<対称性が複数ある場合>
%本研究では，結果的に保存則が1つしかない場合のみを扱った．
%保存則が複数ある場合には，多様体$S$の次元が保存則の数と同じだけ高次元となる．
%この場合，式1が直交する複数の解を持つことになる．
%この場合も理論上は本提案手法で対応可能であるが，回帰する多項式（Eq.()）の組み合わせ数は爆発的に増加し，かつ，ヤコビアンが正則でなくなる確率も増える．
%このため，より効率的な無限小変換の推定が必要になると考えられる．
%例えば，提案手法では，リー群がなす多様体を全空間にわたって回帰した．
%一方で，最終的に推定したい無限小変換は，単位元周辺での接空間である．
%無限のサンプルがあれば，単位元周辺のサンプリング結果を直交基底分解をすれば，この情報は抽出可能である．
%一方で有限のサンプルからこれを行うことは困難である．
%直交基底分解に際して，各種制約条件を導入することで，この問題が回避されることが期待される．
The present study deals only with the case of a single conservation law. If there are multiple conservation laws, the dimension of the manifold $D_a$ becomes increase. In such a case, Eq.~\eqref{diff_eq} has multiple orthogonal solutions. Theoretically, the proposed method can still handle such a problem, but the number of combinations of polynomial regressions [Eq.~\eqref{implicit}] increases exponentially, and the Jacobian matrix is more likely to be singular. Therefore, it is necessary to develop a more efficient method of estimating an infinitesimal transformation. To estimate an infinitesimal transformation, one needs only to estimate a tangent space around the identity element. As there is a finite sample, in the proposed method, the manifold formed by Lie groups is regressed over the entire space. It is expected that the tangent space can be directly estimated by orthogonal basis decomposition by introducing various constraints.\par

%<deep autoencoder以外のモデルによる対称性抽出の可能性について>
%本研究では，時系列データ多様体のモデルとしてdeep autoencoderを用いた．
%一方で，本提案手法でモデルに求められる条件は，多様体上にいるか，多様体の外にいるかを判定可能な写像関数を持つことだけである．
%例えば，GANのdiscriminatorなどは，Trueが多様体上にいることを，Falseが多様体の外にいることを判定可能であることが期待される．また，通常の教師ありDCNNも，中間層から入力を復元するネットワークを追加することで，本提案手法のDNNと置き換え可能である．
%また，DNN以外でもSVLなどの他のモデルの使用も検討可能である．
%重要なことは，本提案手法によって，このような多様なモデルの捉えた物理的情報を抽出可能になったということであり，他の研究者がこれまでに訓練したモデルを活用することで，そこから物理的に有用な情報が抽出されることが期待される．
In this study, we used the deep autoencoder to model the time-series data manifolds; nonetheless, there is no need to use the deep autoencoder. The only requirement for a machine learning model is that it has a mapping function that can determine whether it is on or outside the manifold. From this perspective, the deep autoencoder can be replaced with another type of DNN model, such as a variational autoencoder~\cite{kingma2013auto} or a generative adversarial network~\cite{goodfellow2014generative}. 
%A GAN, like a deep autoencoder, aims to generate data identical to the input data distribution. A DNN called “discriminator” can be constructed, separate from the generative model; this discriminator is trained to determine whether or not generative data are generated from the same distribution of trained data. (The generator model, on the other hand, is trained to generate data wherein the discriminator has the same properties as the training data distribution.) 
%The machine-learning literature hypothesizes about the relationship between dataset and tasks; this gives rise to the “manifold hypothesis,” in which points of different classes are likely to concentrate along various submanifolds, separated by low-density regions of the input space. From the perspective of the manifold hypothesis, it is expected that the GAN discriminator is merely an assessor of whether the input is on or off the manifold. In other words, a GAN can be incorporated into the proposed framework.  
%また，通常のfeed-forward型のDNNに対しても，出力層から入力を復元するネットワークを追加学習することで，本提案手法のDNNと置き換え可能である．
%カーネル法などの他の多様体を埋め込む写像関数を持つ機械学習モデルでも同様のことは実現可能であるはずである．
%このように，本提案枠組みは，機械学習をブラックボックスとして用いた物理データ分析の莫大な既存研究から，物理的な知見を抽出できる可能性をもつ．
Additionally, a feed-forward-type DNN, which is widely used in DNN research, can be used in our proposed method by additionally training a neural network that reconstructs the input data from the output layer of the feed forward neural network. The same method should be feasible for use with machine learning models that have mapping functions that embed data manifolds into the output space (e.g., the kernel method). Thus, the proposed framework can potentially extract interpretable physical knowledge from the wide range of machine learning models. 
Note that the structure of the extracted manifold changes depending on the DNN model and its training settings. 
This is because the reduced model acquired inside the DNN changes depending on the DNN model and the training settings. How to learn time-series dataset using a certain DNN model and training settings are understood as the implicit construction of the reduced model.\par
%<夢>
%本研究によって，力学系時系列データから，explicitな保存量推定が可能であることが示された．これに基づき，これまでに得られた機械学習による物理データ分析と物理学者の縮約モデル構築の試みが融合することが期待される．
In this study, we showed that the proposed framework can infer the hidden conservation laws of a complex system from DNNs that have been trained with physical data of the system. 
On the basis of the obtained results, it is expected that the knowledge of physical data embedded in the trained DNNs in previous studies and the knowledge of physicists can be merged. This should accelerate the research on the construction of reduced models. 

\begin{acknowledgements}
I would like to thank Dr. Y. Ando, Professor S. Goto, Dr. S. Takabe, Professor H. Hino, Professor K. Fukumizu, Professor K. Hukushima, Professor T. Ikegami, Professor K. Ishikawa, Mr. H. Yamashita, Professor Y. Yue, and Mr. K. Sakamoto for useful discussions. This work was supported by KAKENHI grant numbers JP17H01793 and JP19K12111. 
\end{acknowledgements}

\appendix
\section{Derivation of equivalent condition to make Hamiltonian invariant}
\label{appendix_hinv}
The identity condition $H(\textit{\textbf{q}},\textit{\textbf{p}}) \equiv H'(\textit{\textbf{q}},\textit{\textbf{p}})$ has the equivalent expression
\begin{eqnarray}
\forall{(\textit{\textbf{q}},\textit{\textbf{p}})}, H(\textit{\textbf{q}},\textit{\textbf{p}}) = H'(\textit{\textbf{q}},\textit{\textbf{p}}).\label{proof_eq1}
\end{eqnarray}
This condition can be transformed to an equivalent conditional expression represented by a set,
\begin{eqnarray}
\forall{E},\: \{\textit{\textbf{q}},\textit{\textbf{p}}\:\mid \: H(\textit{\textbf{q}},\textit{\textbf{p}}) = E\}= \{\textit{\textbf{q}},\textit{\textbf{p}}\:\mid \:
H'(\textit{\textbf{q}},\textit{\textbf{p}}) = E\},\label{proof_eq2}
%&\Leftrightarrow& 
%\forall{E},\: \{\textit{\textbf{q}},\textit{\textbf{p}}\:\mid \: H(\textit{\textbf{q}},\textit{\textbf{p}}) = E\}= \{\textit{\textbf{Q}},\textit{\textbf{P}}\:\mid \:H'(\textit{\textbf{Q}},\textit{\textbf{P}}) = E\}\label{proof_eq2-2}
\end{eqnarray}
which is proved in Appendix~\ref{appendix1}. Replacing $\textit{\textbf{q}},\textit{\textbf{p}}$ with the transformed parameters $\textit{\textbf{Q}},\textit{\textbf{P}}$ does not change the set: 
$\{\textit{\textbf{q}},\textit{\textbf{p}}\:\mid \: H'(\textit{\textbf{q}},\textit{\textbf{p}}) = E\}= \: \{\textit{\textbf{Q}},\textit{\textbf{P}}\:\mid \: H'(\textit{\textbf{Q}},\textit{\textbf{P}}) = E\}$. Therefore, Eq.~\eqref{proof_eq2} is rewritten as
\begin{eqnarray}
\forall{E},\: \{\textit{\textbf{q}},\textit{\textbf{p}}\:\mid \: H(\textit{\textbf{q}},\textit{\textbf{p}}) = E\}= \{\textit{\textbf{Q}},\textit{\textbf{P}}\:\mid \:
H'(\textit{\textbf{Q}},\textit{\textbf{P}}) = E\}.
\label{proof_eq2-2}
\end{eqnarray}
From the definition of the transformed Hamiltonian $H'$, $H'(\textit{\textbf{Q}},\textit{\textbf{P}}) \coloneqq H\left(\textit{\textbf{q}}(\textit{\textbf{Q}},\textit{\textbf{P}}),\textit{\textbf{p}}(\textit{\textbf{Q}},\textit{\textbf{P}})\right) = H(\textit{\textbf{q}},\textit{\textbf{p}})$ are satisfied. 
By substituting these into Eq.~\eqref{proof_eq2-2}, we obtain the target condition equivalent to the identity condition $H(\textit{\textbf{q}},\textit{\textbf{p}}) \equiv H'(\textit{\textbf{q}},\textit{\textbf{p}})$ as
\begin{eqnarray}
\forall{E},\: \{\textit{\textbf{q}},\textit{\textbf{p}}\:\mid \: H(\textit{\textbf{q}},\textit{\textbf{p}}) = E\}= \{\textit{\textbf{Q}},\textit{\textbf{P}}\:\mid \:
H(\textit{\textbf{q}},\textit{\textbf{p}}) = E\}.
\end{eqnarray}

\section{Derivation of equivalent condition to make canonical equations invariant}
\label{appendix_b}

The identity condition in Eq.~\eqref{proof_motion1},
\begin{eqnarray}
&\:&\textit{\textbf{u}}(\textit{\textbf{q}}_{t},\textit{\textbf{p}}_{t}) \equiv  \textit{\textbf{u}}'(\textit{\textbf{q}}_{t},\textit{\textbf{p}}_{t}) \:\:\:\land\:\:\: 
 \textit{\textbf{v}}(\textit{\textbf{q}}_{t},\textit{\textbf{p}}_{t}) \equiv 
 \textit{\textbf{v}}'(\textit{\textbf{q}}_{t},\textit{\textbf{p}}_{t}),\:\:\:
\end{eqnarray}
has the equivalent expression
\begin{eqnarray}
\forall{(\textit{\textbf{q}}_{t},\textit{\textbf{p}}_{t})},  \:(\textit{\textbf{u}}(\textit{\textbf{q}}_{t},\textit{\textbf{p}}_{t}), \textit{\textbf{v}}(\textit{\textbf{q}}_{t},\textit{\textbf{p}}_{t})) =  (\textit{\textbf{u}}'(\textit{\textbf{q}}_{t},\textit{\textbf{p}}_{t}),\textit{\textbf{v}}'(\textit{\textbf{q}}_{t},\textit{\textbf{p}}_{t})).\label{proof_motion3-0}
\end{eqnarray}
This condition can be transformed to the following equivalent conditional expression represented by a set:
\begin{eqnarray}
&\:&\forall{(\textit{\textbf{q}}_{t+\Delta t},\textit{\textbf{p}}_{t+\Delta t})}, \{\textit{\textbf{q}}_{t},\textit{\textbf{p}}_{t}\:\mid \: (\textit{\textbf{q}}_{t+\Delta t},\textit{\textbf{p}}_{t+\Delta t}) = (\textit{\textbf{u}}(\textit{\textbf{q}}_{t},\textit{\textbf{p}}_{t}),\textit{\textbf{v}}(\textit{\textbf{q}}_{t},\textit{\textbf{p}}_{t}))\}\nonumber\\
&\:&\:=\{\textit{\textbf{q}}_{t},\textit{\textbf{p}}_{t}\:\mid \: (\textit{\textbf{q}}_{t+\Delta t},\textit{\textbf{p}}_{t+\Delta t}) = (\textit{\textbf{u}'}(\textit{\textbf{q}}_{t},\textit{\textbf{p}}_{t}),\textit{\textbf{v}'}(\textit{\textbf{q}}_{t},\textit{\textbf{p}}_{t}))\}.\label{proof_motion3}
\end{eqnarray}
The proof of the equivalence of Eqs.~\eqref{proof_motion3-0} and \eqref{proof_motion3} is a multivariable case of the proof described in Appendix~\ref{appendix1}. 
By treating $\textit{\textbf{q}}_{t+\Delta t},\:\textit{\textbf{p}}_{t+\Delta t}$ as a set of elements, we transform the condition in Eq.~\eqref{proof_motion3} to the equivalent condition (see the proof in Appendix~\ref{appendix2})
\begin{eqnarray}
&\:&\{\textit{\textbf{q}}_{t+\Delta t},\textit{\textbf{p}}_{t+\Delta t},\textit{\textbf{q}}_{t},\textit{\textbf{p}}_{t}\:\mid \: (\textit{\textbf{q}}_{t+\Delta t},\textit{\textbf{p}}_{t+\Delta t}) = (\textit{\textbf{u}}(\textit{\textbf{q}}_{t},\textit{\textbf{p}}_{t}),\textit{\textbf{v}}(\textit{\textbf{q}}_{t},\textit{\textbf{p}}_{t}))\}\nonumber\\
&\:&\:= \{\textit{\textbf{q}}_{t+\Delta t},\textit{\textbf{p}}_{t+\Delta t},\textit{\textbf{q}}_{t},\textit{\textbf{p}}_{t}\:\mid \: (\textit{\textbf{q}}_{t+\Delta t},\textit{\textbf{p}}_{t+\Delta t}) = (\textit{\textbf{u}'}(\textit{\textbf{q}}_{t},\textit{\textbf{p}}_{t}),\textit{\textbf{v}'}(\textit{\textbf{q}}_{t},\textit{\textbf{p}}_{t}))\}.\label{proof_motion4}
\end{eqnarray}
Replacing $\textit{\textbf{q}}_t,\textit{\textbf{p}}_t,\textit{\textbf{q}}_{t+\Delta t},\textit{\textbf{p}}_{t+\Delta t}$ with the transformed parameters $\textit{\textbf{Q}}_{T},\textit{\textbf{P}}_{T},\textit{\textbf{Q}}_{T+\Delta T},\textit{\textbf{P}}_{T+\Delta T}$ does not change the set: 
\begin{eqnarray}
&\:&\{\textit{\textbf{q}}_{t+\Delta t},\textit{\textbf{p}}_{t+\Delta t},\textit{\textbf{q}}_{t},\textit{\textbf{p}}_{t}\:\mid \: (\textit{\textbf{q}}_{t+\Delta t},\textit{\textbf{p}}_{t+\Delta t}) = (\textit{\textbf{u}'}(\textit{\textbf{q}}_{t},\textit{\textbf{p}}_{t}),\textit{\textbf{v}'}(\textit{\textbf{q}}_{t},\textit{\textbf{p}}_{t}))\}\\
&=& \{\textit{\textbf{Q}}_{T+\Delta T},\textit{\textbf{P}}_{T+\Delta T},\textit{\textbf{Q}}_{T},\textit{\textbf{P}}_{T}\:\mid \: (\textit{\textbf{Q}}_{T+\Delta T},\textit{\textbf{P}}_{T+\Delta T}) = (\textit{\textbf{u}'}(\textit{\textbf{Q}}_{T},\textit{\textbf{P}}_{T}),\textit{\textbf{v}'}(\textit{\textbf{Q}}_{T},\textit{\textbf{P}}_{T}))\}.
\end{eqnarray}
Therefore, Eq.~\eqref{proof_motion4} is rewritten as 
\begin{eqnarray}
&\:&\{\textit{\textbf{q}}_{t+\Delta t},\textit{\textbf{p}}_{t+\Delta t},\textit{\textbf{q}}_{t},\textit{\textbf{p}}_{t}\:\mid \: (\textit{\textbf{q}}_{t+\Delta t},\textit{\textbf{p}}_{t+\Delta t}) = (\textit{\textbf{u}}(\textit{\textbf{q}}_{t},\textit{\textbf{p}}_{t}),\textit{\textbf{v}}(\textit{\textbf{q}}_{t},\textit{\textbf{p}}_{t}))\}\nonumber\\
&\:&\:= \{\textit{\textbf{Q}}_{T+\Delta T},\textit{\textbf{P}}_{T+\Delta T},\textit{\textbf{Q}}_{T},\textit{\textbf{P}}_{T}\:\mid \: (\textit{\textbf{Q}}_{T+\Delta T},\textit{\textbf{P}}_{T+\Delta T}) = (\textit{\textbf{u}'}(\textit{\textbf{Q}}_{T},\textit{\textbf{P}}_{T}),\textit{\textbf{v}'}(\textit{\textbf{Q}}_{T},\textit{\textbf{P}}_{T}))\}.
\label{proof5}
\end{eqnarray}
From the definition of the transformed canonical equations [Eqs.~\eqref{canf1} and \eqref{canf2}], we obtain 
\begin{eqnarray}
%u'\left(\textit{\textbf{Q}}_T,\textit{\textbf{P}}_T\right) &=& u'\left(\textit{\textbf{Q}}_T(\textit{\textbf{q}}_t,\textit{\textbf{p}}_t, \theta),\textit{\textbf{P}}_T(\textit{\textbf{q}}_t,\textit{\textbf{p}}_t, \theta)\right) = u(\textit{\textbf{q}}_t,\textit{\textbf{p}}_t),\\ v'\left(\textit{\textbf{Q}}_T,\textit{\textbf{P}}_T\right) &=& v'\left(\textit{\textbf{Q}}_T(\textit{\textbf{q}}_t,\textit{\textbf{p}}_t, \theta),\textit{\textbf{P}}_T(\textit{\textbf{q}}_t,\textit{\textbf{p}}_t, \theta)\right) = v(\textit{\textbf{q}}_t,\textit{\textbf{p}}_t).
\left(\textit{\textbf{Q}}_{T+\Delta T},\textit{\textbf{P}}_{T+\Delta T}\right) &=& \left(\textit{\textbf{u}'}(\textit{\textbf{Q}}_{T},\textit{\textbf{P}}_{T}),\textit{\textbf{v}'}(\textit{\textbf{Q}}_{T},\textit{\textbf{P}}_{T})\right)\\
\Leftrightarrow\left(\textit{\textbf{q}}_{t+\Delta t},\textit{\textbf{p}}_{t+\Delta t}\right) &=& \left(\textit{\textbf{u}}(\textit{\textbf{q}}_{t},\textit{\textbf{p}}_{t}),\textit{\textbf{v}}(\textit{\textbf{q}}_{t},\textit{\textbf{p}}_{t})\right).
\end{eqnarray}
By substituting this into Eq.~\eqref{proof5}, we obtain the target condition equivalent to the identity condition in Eq.~\eqref{proof_motion1} as
\begin{eqnarray}
&\:&
\{\textit{\textbf{q}}_{t+\Delta t},\textit{\textbf{p}}_{t+\Delta t},\textit{\textbf{q}}_{t},\textit{\textbf{p}}_{t}\:\mid \: (\textit{\textbf{q}}_{t+\Delta t},\textit{\textbf{p}}_{t+\Delta t}) = (\textit{\textbf{u}}(\textit{\textbf{q}}_{t},\textit{\textbf{p}}_{t}),\textit{\textbf{v}}(\textit{\textbf{q}}_{t},\textit{\textbf{p}}_{t}))\}\nonumber\\
&\:&\:= \{\textit{\textbf{Q}}_{T+\Delta T},\textit{\textbf{P}}_{T+\Delta T},\textit{\textbf{Q}}_{T},\textit{\textbf{P}}_{T}\:\mid \: (\textit{\textbf{q}}_{t+\Delta t},\textit{\textbf{p}}_{t+\Delta t}) = (\textit{\textbf{u}}(\textit{\textbf{q}}_{t},\textit{\textbf{p}}_{t}),\textit{\textbf{v}}(\textit{\textbf{q}}_{t},\textit{\textbf{p}}_{t}))\}.
\end{eqnarray}

\section{Proof of Eq.~\eqref{proof_eq1}$\Leftrightarrow$Eq.~\eqref{proof_eq2} and Eq.~\eqref{proof_motion3-0}$\Leftrightarrow$Eq.~\eqref{proof_motion3}}
\label{appendix1}
The problem can be abstracted as the proposition below: 
\begin{eqnarray}
\forall x, f(x) = g(x) \Leftrightarrow \forall E, \{x | f(x) = E\} = \{x | g(x) = E\},
\end{eqnarray}
where $f(x)$ and $g(x)$ are single-valued functions. \par
\begin{eqnarray}
\label{proofc1}
\bullet\:\:{\rm Proof\:\:of\:\:} \forall x, f(x) = g(x) \rightarrow  \forall E, \{x | f(x) = E\} = \{x | g(x) = E\} 
\end{eqnarray}
The contrapositive of \eqref{proofc1} is $\exists E, \{x | f(x) = E\} \neq \{x | g(x) = E\} \rightarrow  \exists x, f(x) \neq g(x) $. This contrapositive is proved as follows. Since $ \exists E, \{x | f(x) = E\} \neq \{x | g(x) = E\} $, there exists $E'$ and $x'$, which satisfy $f(x')=E'$, but $g(x') \neq E'$. 
Therefore, $\exists x, f(x) \neq g(x) $ is satisfied because $f(x') \neq g(x')$.\par
\begin{flushright}
$\Box$
\end{flushright}\par
\begin{eqnarray}
\label{proofc2}
\bullet\:\:{\rm Proof\:\:of\:\:} \forall E, \{x | f(x) = E\} = \{x | g(x) = E\} \rightarrow \forall x, f(x) = g(x) 
\end{eqnarray}
The contrapositive of \eqref{proofc2} is $\exists x, f(x) \neq g(x) \rightarrow  \exists E, \{x | f(x) = E\} \neq \{x | g(x) = E\}$. This contrapositive is proved as follows. Select one $x'$ from $x$, which satisfies $f(x') \neq g(x')$ and $f(x')=E'$. Since $f(x)$ is a single-valued function, $x'$ is not included in the set of $x$ that satisfies $g(x)=E'$. Thus, $\{x | f(x) = E'\} \neq \{x | g(x) = E'\} $ holds. Therefore, $\exists E, \{x | f(x) = E\} \neq \{x | g(x) = E\}$ is satisfied.\par
\begin{flushright}
$\Box$
\end{flushright}\par

\section{Proof of Eq.~\eqref{proof_motion3}$\Leftrightarrow$Eq.~\eqref{proof_motion4}}
\label{appendix2}
The problem can be abstracted as the proposition below: 
\begin{eqnarray}
\forall b, \{x | \:f(x) = b\} = \{x | \:g(x) = b\}\Leftrightarrow \{x,b | \:f(x) = b\} = \{x,b | \:g(x) = b\},
\end{eqnarray}
where $f(x)$ and $g(x)$ are single-valued functions.\par
\begin{eqnarray}
\label{proofd1}
\bullet\:\:{\rm Proof\:\:of\:\:} \forall b, \{x | \:f(x) = b\} = \{x | \:g(x) = b\}\rightarrow \{x,b | \:f(x) = b\} = \{x,b | \:g(x) = b\}
\end{eqnarray}
The contrapositive of \eqref{proofd1} is $\{x,b | \:f(x) = b\} \neq \{x,b | \:g(x) = b\} \rightarrow \exists b, \{x | \:f(x) = b\} \neq \{x | \:g(x) = b\}$. This contrapositive is proved as follows. Since $\{x,b | \:f(x) = b\} \neq \{x,b | \:g(x) = b\}$, there is a set of $x'$ and $b'$, which satisfies $f(x')=b'$ and $g(x')\neq b'$. Therefore, $\{x | \:f(x) = b'\} \neq \{x | \:g(x) = b'\}$ holds. It means that $\exists b, \{x | \:f(x) = b\} \neq \{x | \:g(x) = b\}$ is satisfied.\par
\begin{flushright}
$\Box$
\end{flushright}\par
\begin{eqnarray}
\label{proofd2}
\bullet\:\:{\rm Proof\:\:of\:\:} \{x,b | \:f(x) = b\} = \{x,b | \:g(x) = b\} \rightarrow \forall b, \{x | \:f(x) = b\} = \{x | \:g(x) = b\}
\end{eqnarray}
The contrapositive of \eqref{proofd2} is $\exists b, \{x | \:f(x) = b\} \neq \{x | \:g(x) = b\} \rightarrow \{x,b | \:f(x) = b\} \neq \{x,b | \:g(x) = b\}$. This contrapositive is proved as follows. Since $\exists b, \{x | \:f(x) = b\} \neq \{x | \:g(x) = b\}$, there is a set of $b'$ and $x'$, which satisfies $f(x')=b'$ and $g(x')\neq b'$. Therefore, $\{x,b | \:f(x) = b\} \neq \{x,b | \:g(x) = b\}$ is satisfied.\par
\begin{flushright}
$\Box$
\end{flushright}\par

\section{Replica exchange Monte Carlo (REMC) method and its parameters}
\label{remc_params}
Using $A' \coloneqq (a_{11},a_{12},a_{21},\cdots,a_{2d\:2d})$, we re-express Eq.~\eqref{samp_func0} as
\begin{equation}
{\rm P}(A') = \frac{1}{Z} \exp\left[-\frac{N}{2\sigma^2} E_{\rm samp}(A')\right]. 
\end{equation}
The REMC method takes samples from the joint density 
  \begin{equation}
{\rm  P}(A'^1,\cdots,A'^l,\cdots, A'^L)=\prod_{l=1}^L\frac{1}{Z} \exp\left[-\frac{N}{2\sigma_l^2} E_{\rm samp}(A')\right],
\end{equation}
where $\sigma_l>\sigma_{l+1}$ and $\sigma_L = \sigma$. 
In the REMC method, sampling from the joint density ${\rm  P}(A'^1,\cdots,A'^l,\cdots, A'^L)$ is performed on the basis of the following updates.
 \\

\begin{itemize}
\item[{\bf 1}]{\bf Sampling from each density ${\rm  P}(A'^1,\cdots,A'^l,\cdots, A'^L)$}\par
\parindent=3pt Sampling $A'^l$ from ${\rm P}(A'^l) \coloneqq \frac{1}{Z_l} \exp\left[-\frac{N}{2\sigma_l^2} E_{\rm samp}(A'^l)\right]$, where $Z'$ is the normalization constant.  The sampling is performed by a conventional Monte Carlo method, such as the Metropolis\textendash Hastings algorithm~\cite{hastings1970monte}. 

 \item[{\bf 2}] {\bf Exchange between two densities corresponding to noise intensity $\sigma$}\par
 \parindent=3pt The exchanges between the configurations $A'^l$ and 
 $A'^{l+1}$ correspond to adjacent inverse temperatures following the probability $R=\min(1,r)$, where\par
  
\begin{equation}
\begin{split}
\nonumber
r &= \frac{{\rm  P}(A'^1,\cdots, A'^{l+1}, A'^l,\cdots,A'^L)}{{\rm  P}(A'^1,\cdots, A'^l, A'^{l+1},\cdots,A'^L)}\\
&=\frac{{\rm  P}(A'^{l+1}){\rm  P}(A'^l)}{{\rm  P}(A'^l){\rm  P}(A'^{l+1})}\\
&=\exp\left\{\frac{N}{2}[\sigma_{l+1}^{-2} - \sigma_{l}^{-2}][E(A'^{l+1}) - E_{\rm samp}(A'^l)]\right\}.
\end{split}
\end{equation}

\end{itemize}
Sampling from a distribution with a larger $\sigma_l$ tends not to have a local minimum. 
%On the other hand, sampling from a distribution with 
%a larger $\beta$ corresponds to sampling from a distribution with 
%local minima. 
Hence, sampling from the joint density ${\rm  P}(A'^1,A'^2\cdots A'^L)$ overcomes the local minima in distributions with small $\sigma_l$ and enables the rapid convergence of sampling.\par
In the execution of EMC sampling, we adopted the Metropolis\textendash Hastings algorithm~\cite{hastings1970monte} to sample each state of $\sigma_l$. 
When we performed the Metropolis\textendash Hastings sampling, a candidate for the next sample $a_{ij}^{l\:\:\rm next}$ is picked from the conditional probability distribution with precondition $a_{ij}^{l\:\:\rm previous}$
\begin{equation}
{\rm P}(a_{ij}^{l\:\:\rm next} | a_{ij}^{l\:\:\rm previous}) = \frac{1}{2U_l}\:\:\:(-U_l \leq a_{ij}^{l\:\:\rm next} \leq U_l),
\end{equation}
where $U_l$ is set as
\begin{eqnarray}
\label{mhc}
U_l = \left\{ \begin{array}{ll}
C & (eN\sigma_l^{-2}\geq 1) \\
\frac{C}{(eN\sigma_l^{-2})^{d}} & (eN\sigma_l^{-2} < 1) \\
\end{array}. \right.
\end{eqnarray}
$C$, $d$, and $e$ in Eq.~\eqref{mhc} are set as Table~\ref{tbl2} for the evaluation of the proposed method in Sec. \ref{results}. In case~(\rnum{2}) constant-velocity linear-motion, the sampling parameters $C$ of $a$ and $b$ in Eq.~\eqref{sampab} were set as different values [The values are described as ($C$ of $a$) / ($C$ of $b$) in the column of (\rnum{2}) constant-velocity of Table~\ref{tbl2}].  
%また，EMC法によるサンプリングは$L=40$の温度準位を用意した上で，
%サンプリングの最初の10,000 stepをburn inとして棄却した後に，
%100,000 step分のサンプリング結果を用いた．
Each state of $\sigma_l$ was determined following the exponential function~\cite{nagata2008asymptotic}:
\begin{eqnarray}
\sigma_l^{-2} = \left\{ \begin{array}{ll}
0.0 & (l=1) \\
\sigma_{\rm min}^{-2} \gamma^{(l-1)-L} & (l\neq1) \\
\end{array} ,\right.
\end{eqnarray}
where $\sigma_{\rm min}$ is set as root-mean-square error (RMSE) for $A = \boldsymbol{I}$ in trained DNN, because it represents the minimum value of $E_{\rm samp}$. 
$L$ and $\gamma$ are set as shown in Table~\ref{tbl2} for each case. 

\begin{table}[t]%The best place to locate the table environment is directly after its first reference in text
\caption{\label{tbl2}%
Parameters of REMC method.
}
\begin{ruledtabular}
\begin{tabular}{c|cccc}
\textrm{Parameter name}&
\textrm{(\rnum{1}) Half sphere}&
\textrm{(\rnum{2}) Constant-velocity}&
\textrm{(\rnum{3}) Central force}&
\textrm{(\rnum{4}) Collective motion}\\
\colrule
Sampling size $N_a$ & 10,000 & 10,000 & 10,000 & 10,000\\
L & 20 & 30 & 30 & 30\\
$\gamma$ & 1.4 & 1.9 & 1.4 & 1.4\\
$C$ & 3.0 & 0.03 / 0.3 & 0.3 & 0.3\\
$d$ & 0.6 & 0.7 & 0.8 & 0.8\\
$e$ & 5.0 & 1.0 & 5.0 & 5.0\\
$\sigma_{\rm min}$ & 4.42 $\times$ $10^{-2}$ & 5.41 $\times$ $10^{-5}$ & 1.67 $\times$ $10^{-1}$ & 8.0\\
Burn-in length & 1.000 & 10,000 & 1,000 & 10,000\\
Selected noise intensity $\sigma_{\rm noise}$ & 8.66 $\times$ $10^{-2}$ & 5.41 $\times$ $10^{-5}$ & 3.45 & 71.3\\
\end{tabular}
\end{ruledtabular}
\end{table}

\section{Noise intensity of sampling}
\label{appendix_sec3}
\begin{figure}
 \begin{center}
  \includegraphics[width=10.0cm]{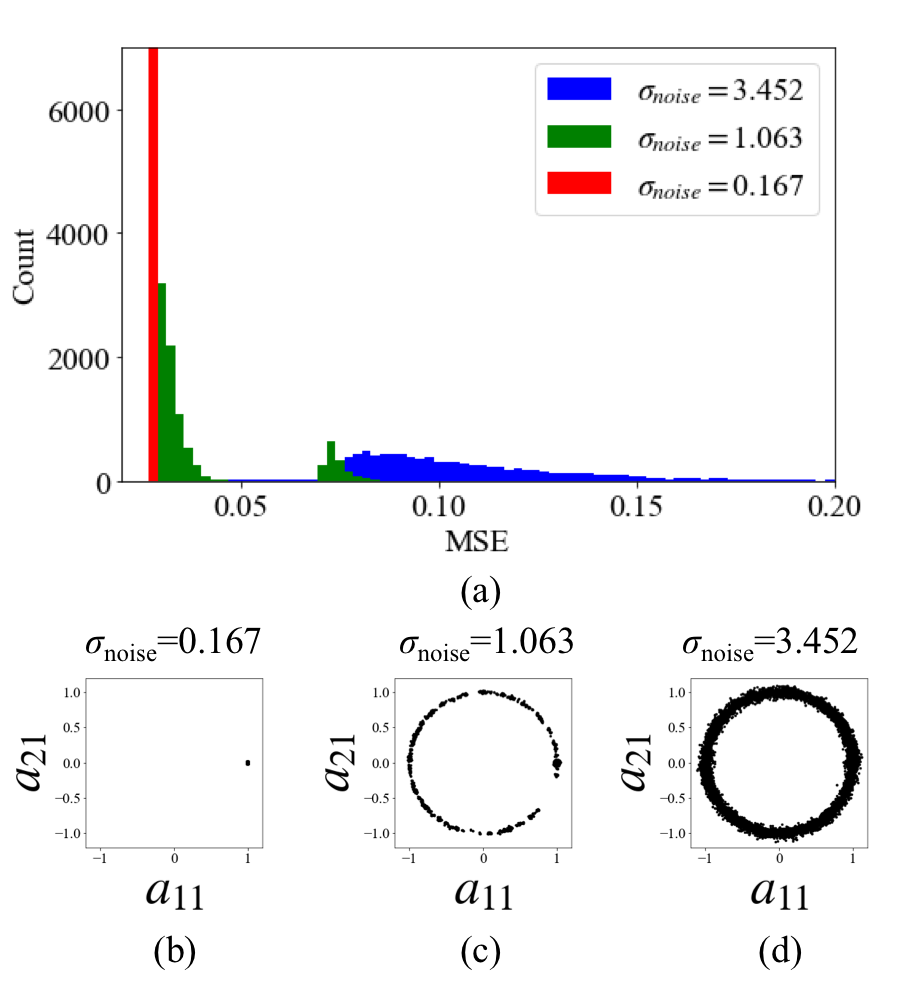}
  \caption{Qualitative transition of sampling results due to the increase in noise intensity. 
The figures describe the qualitative transition of case~(\rnum{3}) central force system where rotation symmetry exists. 
(a) Distributions of MSE with different noise intensities. 
(b), (c), and (d) Sampling results of $a_{11}$ and $a_{21}$ at each noise intensity. }
  \label{fig_appendix1}
 \end{center}
\end{figure}
%$\sigma$の違いによって，低MSEにピークを持つサンプリング結果と，高MSEに対応するサンプリング結果が得られる．
Depending on the difference in $\sigma_{\rm noise}$, the sampling results corresponding to low MSE and the sampling results corresponding to high MSE are obtained [Fig.~\ref{fig_appendix1}(a)]. 
%低MSE領域では，単位行列に対応する変換行列がサンプリングされている．
In the low-MSE region, the transformation matrix corresponding to the identity matrix is sampled [Fig.~\ref{fig_appendix1}(b)]. 
%高MSE領域では，回転行列に対応する変換行列がサンプリングされている．
In the high-MSE region, the transformation matrix corresponding to the rotation matrix is sampled [Fig.~\ref{fig_appendix1}(d)]. 
%中間ノイズ強度では，両方の中間のサンプリングが実現される．
At intermediate noise intensities, sampling between both conditions is achieved [Fig.~\ref{fig_appendix1}(c)]. 
%従って，より大域的な変換のピークのみが実現されるノイズ強度を選択すれば，得たい対象変換が得られる．
On the basis of such a structure, in this research, we select the noise intensity $\sigma_{\rm noise}$ that realizes the non-identity transformation such as $\sigma_{\rm noise} = 3.452$ of Fig.~\ref{fig_appendix1}. For the evaluation of the proposed method in Sec. \ref{results}, we select the noise intensity $\sigma_{\rm noise}$ for each evaluation case as shown in Table~\ref{tbl2}.\par

\section{Estimation of likelihood for the model selection}
\label{est_likeli}
Under the assumption that $N_a$ samples of transformation are given with Gaussian noise, the following likelihood is defined for a statistical selection of implicit function. 
\begin{eqnarray}
\label{likelihood}
P\left(\vec{b}_1,\vec{b}_2,\cdots,\vec{b}_{N_a}\right) &=& \frac{1}{Z}\exp\left\{ -\frac{1}{2\sigma_{b}^2}\sum_{n_a=1}^{N_a} D\left[\vec{b}_{n_a}, f(c_k, b_1,b_2,\cdots,b_{d_{\theta}'};\beta,\gamma,d_{\theta}')\right]^2 \right\},\\
\label{normalizeconstant}
Z &=& \int_{-\infty}^{\infty} d\vec{b}_{n_a} \exp\left\{ -\frac{1}{2\sigma_{b}^2}D\left[\vec{b}_{n_a}, f(c_k, b_1,b_2,\cdots,b_{d_{\theta}'};\beta,\gamma,d_{\theta}')\right]^2 \right\},\\
\sigma_{b} &=& \left\{\frac{1}{N_a} \sum_{n_a=1}^{N_a} D\left[\vec{b}_{n_a}, f(c_k, b_1,b_2,\cdots,b_{d_{\theta}'};\beta,\gamma,d_{\theta}')\right]^2\right\}^{\frac{1}{2}},\\
\vec{b}_{n_a} &=& (c_k, b_1,b_2,\cdots,b_{d_{\theta}'})_{n_a},
\end{eqnarray}
where $D\left[\vec{b}_{n_a}, f(c_k, b_1,b_2,\cdots,b_{d_{\theta}'};\beta,\gamma,d_{\theta}')\right]$ is the minimum distance from a data point $\vec{b}_{n_a}$ to a subspace defined by the implicit function $f(c_k, b_1,b_2,\cdots,b_{d_{\theta}'};\beta,\gamma,d_{\theta}') = 0$. 
The normalized constant $Z$ is estimated numerically as the Riemann sum.

\section{DNN model and its training parameters}
\label{dnn_params}
\begin{table}[t]
\caption{Parameters of DNN and its training. In the ``Network structure'', the number of nodes is shown in the order from left to right: input layer -- first layer -- second layer -- third layer -- output layer.}
\label{tbl_dnnsettings}
\begin{ruledtabular}
\begin{tabular}{c|cccc}
\textrm{Parameter name}&
\textrm{(\rnum{1}) Half sphere}&
\textrm{(\rnum{2}) Constant-velocity}&
\textrm{(\rnum{3}) Central force}&
\textrm{(\rnum{4}) Collective motion}\\
\colrule
Training datasize $N$ & 1,671 & 1,000 & 1,000 & 5,000\\
Network structure & 3-10-2-10-3 & 4-10-1-10-4 & 8-20-1-20-8 & 8-20-1-20-8\\
Activation function & sigmoid & tanh & sigmoid & sigmoid\\
Training algorithm & Adam & Adam & Adam & Adam\\
%Learning rate & 0.0001 & 0.00001 & 0.0001 & 0.0001\\
Training iteration & 50,000 & 100,000 & 50,000 & 50,000\\
Minibatch size & 10 & 30 & 10 & 10\\
Library & theano \cite{Bastien-Theano-2012, bergstra+al:2010-scipy} & scikit-learn \cite{scikit-learn} & theano & theano\\
\end{tabular}
\end{ruledtabular}
\end{table}
%ここでは，力学系の時系列データの学習に用いたDNNとその学習設定について記述する．
In this section, we describe the DNN models and their training settings.\par
%本研究では，DNNモデルとしてdeep auto-encoderを用いた．
In this study, we used deep autoencoders as DNN models. 
%それらは，全てのケースでinpulayerと3層のhidden layerそして，output layerから構成されていた．
In all cases~(\rnum{1}), (\rnum{2}), (\rnum{3}), and (\rnum{4}), the DNNs consisted of an input layer, three hidden layers, and an output layer. 
%各層のノード数は，表1のNetwork structureにあるように設定された．
The number of nodes in each layer was set as shown in the ``Network structure'' in Table~\ref{tbl_dnnsettings}.\par
%表では，左から，input layer - first layer - second layer - third layer - output layerの順番でノード数が表わされている．
%また，daeの活性化関数としては，表1のActivation functionの行にあるような関数が用いられた．
The activation functions of the deep autoencoders were set as the sigmoid or hyperbolic tangent functions as shown in the ``Activation function'' in Table~\ref{tbl_dnnsettings}. 
%ここで，sigmoid関数とは
The sigmoid function is defined as
\begin{equation}
{\rm sigmoid}(x) = \frac{1}{1 + \exp(-x)},
\end{equation}
%と定義され，tanh関数は
and the tanh function is defined as
\begin{equation}
\tanh(x) = \frac{\exp(x) - \exp(-x)}{\exp(x) + \exp(-x)}.
\end{equation}\par
%と定義される．
%本研究のケースb)--d)で学習に用いた時系列データは，単一の初期値からスタートして生成した時系列データを用いた．
%The time-series datasets used for training the DNNs were generated by numerical simulations starting from certain initial states, respectively. 
%一般には単一の初期値からスタートする場合に，$S_i$に到達できない領域が存在する可能性があることに注意する必要がある．
%In general, there may be regions where $S_i$ cannot be reached starting from one initial state. 
%本研究で対象とした系では，到達できない領域が存在しない[ケースb)とc))か，ある初期値から始まって生じる準安定状態を縮約モデル化の対象とした（d)為，問題は生じない．
%As we described in the main text, in all physical cases (b), (c), and (d), there is no linear continuous transformation that transforms part of $S_i$ with one initial state to part of $S_i$ where cannot be reached starting from that initial state. 
%Therefore, in case of (b), (c), and (d), there are always candidates for invariant transformations of $S_i$ in invariant transformations of part of $S_i$ with one initial value. 
%Therefore, we used time-series datasets generated from numerical simulations starting from certain initial states. 
%そのデータから，サンプル数分の時間ステップ分のデータを取り出して学習に用いた．
%From the simulated time-series data, the data up to the maximum time was prepared for the training of DNN. 
%用いたサンプル数は，表1のTrainng datasizeに記載されている．
The numbers of samples used for training DNN are shown in Table~\ref{tbl_dnnsettings} as ``Training datasize $N$''. 
%学習にはadamアルゴリズムが用いられた．
The Adam method~\cite{kingma2014adam} was used for training. 
%また，学習回数は表1のLearning iterationにある回数にそれぞれ設定された．
The training iterations are shown in Table~\ref{tbl_dnnsettings}. 
%また，学習時ではデータを表1のMinibacth sizeにあるような単位に分けて学習を行った．
In the training, the data were divided into minibatches whose sizes are shown in Table~\ref{tbl_dnnsettings} as ``Minibatch size''.
%使用したライブラリについて，表1のLibraryの行にまとめる．
%The libraries used for training the DNNs are shown in the ``Library'' in Table~\ref{tbl_dnnsettings}. 
%使用したライブラリがbのみ違うのは，bは比較的規模の小さなネットワークのため，GPUを使用する必要がなかったためである．
%Only in case of (b), the scikit-learn library was used for training the DNN. 
%Because case (b) used a deep autoencoder with a relatively small node number, there was no need to use a GPU. 
%また，bのみ各種パラメータが特殊なのは，使用したライブラリのdefault値を用いたことが理由である．
%The reason why various parameters are special only in b is that the default value of scikit-learn library was used. 

\bibliography{main}% Produces the bibliography via BibTeX.

\end{document}